\documentclass[onecolumn,article]{revtex4}
\pdfoutput=1
\usepackage[sort&compress]{natbib}
\bibpunct{[}{]}{,}{n}{,}{,}

        \usepackage[usenames,dvipsnames]{color}
        \usepackage[pdftex]{graphicx}
        \usepackage{epstopdf}
        \DeclareGraphicsExtensions{.pdf, .eps, .png}

\begin{document}
%
\title{Multiscale modelling of bionano interface}
\author{Hender Lopez$^1$}

\author{Erik Brandt$^2$}

\author{Alexander Mirzoev$^2$}

\author{Dmitry Zhurkin$^2$}

\author{Alexander Lyubartsev$^2$}

\author{Vladimir Lobaskin$^1$}

\affiliation{$^1$School of Physics, Complex and Adaptive Systems Lab, University College Dublin, Belfield, Dublin 4, Ireland\\
$^2$Department of Materials and Environmental Chemistry, Stockholm University, SE-10691, Stockholm, Sweden}

\maketitle

\section{Introduction} \label{sec::intro}

Over the last decade, in vitro and in vivo experiments have produced significant amount of veritable information that can be integrated into theoretical models with the aim of predicting possible health and environmental effects of engineered nanoparticles (NP) \cite{1-7}. However, even the most systematic studies leave the question of precise toxicity mechanisms associated with nanoparticles wide open \cite{1-3,1-8,1-9}. An important finding arising from these studies is that the toxic effects can emerge either from membrane damage or from interaction of nanoparticles, once they are inside the cell, with the internal cell machinery. Therefore, an evaluation of possible risks should include an assessment of nanoparticle ability to penetrate, modify, or destroy the cell membrane \cite{1-9}. Being selectively permeable, membranes participate in control of the transport of vital substances into and out of cells. Whereas some biomolecules may penetrate or fuse with cell membranes without overt membrane disruption, no synthetic material of comparable size has shown this property \cite{1-10}. Among the factors determining the outcome of NP-membrane interaction the surface properties of nanomaterials play a critical role, which can implicate the membrane or plasma proteins in conditioning NP prior to cell penetration.

The detailed understanding of the crucial stages of NP-cell membrane interaction can be achieved with computer simulation. Molecular dynamics is now a well-recognized tool for studying intermolecular interactions, self-assembly, and structure of biomolecules or their complexes. The reliability and predictive character of molecular modelling has improved significantly during the last few years, with development of new, carefully parameterized force fields, simulation algorithms, and greatly increased computer power \cite{1-11}. The role of computer simulation is now well recognized in many fields e.g. drug design and toxicology \cite{1-12,1-13,1-14}. In the same way, one can attempt to predict the detrimental effect of NPs from physical considerations. Establishing a qualitative and quantitative connection between physicochemical properties of the NP and their effect on biological functioning of membranes can help to identify the possible pathways leading to toxicity and give a mechanistic interpretation of toxicological data. To achieve this goal, one has to understand the processes occurring at bionano interface or on the initial stages of contact between the foreign nanomaterial and the organism such as formation of NP-biomolecule complexes, NP-cell membrane interaction, and NP uptake into the cell.

Understanding the corona formation and NP uptake requires one to address the lengthscales at the range of up to hundred nanometres, which is currently beyond the reach of direct atomistic modelling. Though lipid membranes have been very intensively studied by molecular simulations during last decade \cite{1-54}, in general, modelling NP translocation through a lipid membrane is a significant challenge. Depending on the size of the nanoparticle and any associated proteins (corona) tens of thousands, or more, lipid and other molecules may be needed to model a representative fraction of the membrane. For small (under 5 nm) NPs, cytotoxicity effects such as membrane disruption and poration can be addressed at the atomistic scale and at this scale significant insights have already been gained using molecular simulation using atomistic or coarse-grained force fields \cite{1-58,1-58a,1-58b,1-58c}. To assess interactions of larger NPs with membranes mesoscopic simulations based on greatly reduced number of degrees of freedom are required. To build a quantitative mesoscale model, information on NP-biomolecule association should be transferred from atomistic simulations to the mesoscopic scale.

Many of today's coarse-grained (CG) models use empirical parameterisation of effective interaction potentials. There exist several basic approaches for systematically constructing effective CG potentials from the results of atomistic simulations. One common approach is based on reproduction of forces for specific snapshots of the system (the force matching approach \cite{1-60,1-60a} and the other, is based on fitting of structural properties, for which the radial distribution functions are typically used (the inverse Monte Carlo (IMC) or Newton Inversion method) \cite{1-61,1-62}. The IMC method was used to build coarse-grained models of various molecular systems including ion-DNA solution \cite{1-65}.
In the same spirit, CG models of plasma and membrane proteins have been developed \cite{Toz2005,BerDes2009,Tak2012,WeiKno2013} using the united-atom scheme, i.e. replacing the common groups of atoms by single beads, and thus drastically reducing the number of degrees of freedom. A systematic coarse-graining based on the all-atom presentations will preserve the shape and size of the relevant molecules and thus molecular specificity. In this approach we sacrifice a number of internal degrees of freedom, such as protein conformations, which can be justified a posteriori. Although neglecting the protein internal degrees of freedom is a necessarily shaky approximation, this could be the most beneficial one as we can get around the dynamic bottlenecks related to slow protein unfolding.

Similar to molecules, one can use IMC and other coarse-graining methods to model effective interactions between nanoparticles \cite{1-14,1-64,1-67}. Thus, the construction of the mesoscale modelling tool  involves the following steps, with each consecutive stage based on a systematic coarse-graining of the more detailed description and validated by experimental data:
\begin{figure}[tbh]
\centering
\includegraphics[width=\hsize]{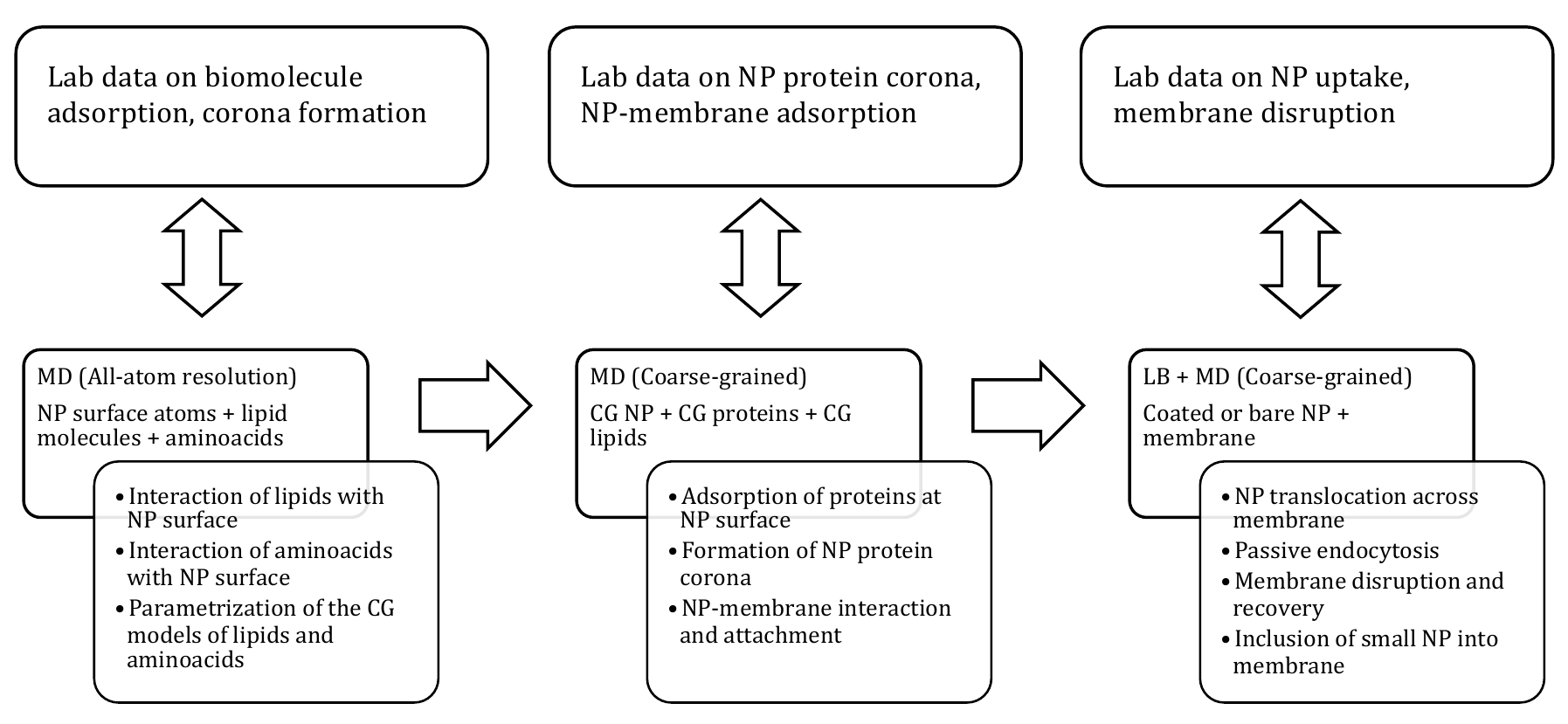}
\caption{Scheme of the multiscale simulation approach for modelling NP uptake.}
\label{fig::diagram}
\end{figure}

In the following sections, we describe a set of CG tools that allow one to simulate the
uptake of the NP-protein corona complex by a lipid bilayer. The remainder of the paper is constructed as follow. First
we report a CG model to calculate the adsorption energies and the most favorable adsorption orientations
of proteins onto a hydrophobic NP.
The proposed method is then used to calculate the adsorption energies of the two common  proteins in human blood
onto NPs with negative or positive surface charge or neutral surface. We also report the effect of the NP radius
on the adsorption energies and validate the proposed methodology against full atomistic simulations.
Then, in Sec.~\ref{sec::bilayer} we describe a methodology in which full atomistic simulations
of a lipid bilayer and various lipid-cholesterol mixtures are used for the extraction CG pair potentials.
We also compare and validate the predictions of simulations at molecular and CG level. In Sec.~\ref{sec::NP-bilayer}, we
present a CG simulation of the interaction a bare NP and of a NP-protein complex with a lipid bilayer.
Finally, in Sec.~\ref{sec::conclusions} we summarise the main results.

\section{Nanoparticle-protein interaction} \label{sec::NP-P}

It is now well accepted that foreign surfaces are modified by the adsorption of biomolecules such as proteins or lipids in a biological environment, and that cellular responses to materials in a biological medium might reflect the adsorbed biomolecule layer, rather than the material itself \cite{1-71}. Recently, the concept of the NP-protein corona has been introduced to describe the proteins in association with NPs in biological fluids \cite{1-72,1-72a,1-72b,1-72c}. The composition of NP corona is flexible and is determined by many affinity constants and concentrations of the components of the blood plasma. One can speculate that in many practically relevant situations, the protein corona is the surface that is exposed to the cell membrane and is the entity the cell protective mechanisms have to deal with. Thus, for most cases it is more likely that the biologically relevant unit is not the particle, but a nanoobject of specified size, shape, and protein corona structure. Naked particle surfaces will have a much greater (non-specific) affinity for the cell surface than a particle hiding behind a corona of ``bystander'' proteins - that is proteins for which no suitable cellular recognition machinery exists. The evidence suggests that, in comparison to typical cell-membrane-biology event timescales the particle corona is likely to be a defining property of the particle in its interactions with the cell surface, whether it activates cellular machinery or not. Similar observations and outcomes exist for particles inside the cell, in key locations, though we cannot discuss details here \cite{1-72,1-72a,1-72b,1-72c}.
	We assume that the actual content of the corona is determined by (i) the NP exposure to the protein solution (blood plasma), (ii) a competition between the adsorbed proteins and the glycoproteins/membrane lipids.
We will model the protein and lipid interaction with the NP surface at the CG united-atom level for selected set of proteins and lipids (see Table \ref{tab::PDB}).
These simulations will provide interaction energies and will be used to predict the kinetics of protein/lipid corona formation. The data on aminoacid interaction with NP will allow us to compute binding affinities of arbitrary proteins of known structure within an additive model implying that the total protein-NP interaction energy is computed as a sum of NP interactions with aminoacids in contact with NP surface. This assumption will be further verified by comparison with results of large scale molecular dynamics simulations for a few selected protein types and mean force potential calculations. From the typical protein concentrations and adsorption energies one can also predict the average content of the corona using ideal adsorbed solution theory \cite{1-85}. It is important to understand that at this stage we would be able neither to scan all the plasma proteins nor to take into account any change of protein conformation or bonding between the adsorbed proteins. However, as the effect of the corona is still largely unknown, we can only hope to capture the most important contributions of the plasma protein to the NP dispersion stability and their interaction with the cell membrane.
\begin{table}
\centering
\begin{tabular}{ c c c c c}
\hline
\hline
Protein & PDB ID & Abbreviation & Weight fraction & Molar mass\\
 &  &  & in plasma, \% & in, kDa\\
\hline
 Human Serum Albumin & 1N5U & HSA & 5.0 & 67 \\
 Fibrinogen & 3GHG & Fib & 0.4 & 340 \\
\hline
\hline
\end{tabular}
\caption{Proteins, PDB ID used for the coarse graining and the abbreviations used
in the text, and their size and abundance in human blood plasma.}
\label{tab::PDB}
\end{table}

Due to complexity of blood plasma, we can only model it at a simplified level. It seems reasonable to include the elements, which are more likely to affect the NP interactions and aggregation, and mediate their interaction with the membrane. The plasma can then be modelled a solution of biomolecules in an implicit solvent with a dielectric constant of water and the Debye length corresponding to physiological ionic strength, van der Waals interactions set to corresponding triplets NP-protein-water, or protein-water-protein, and appropriate surface charges on the molecules. In this work we study the adsorption of two of the most abundant proteins
in blood plasma, Human Serum Albumin (HSA) and Fibrinogen (Fib). In Table \ref{tab::PDB}, we summarize their relative content in blood and their molar mass.
Although this two proteins represent are important components of the blood plasma because of their abundance, recent observations \cite{1-36,1-72,1-72a,1-72b} demonstrate that the protein corona can include hundreds of different plasma proteins. As of now, it is not known which proteins dominate the content of the corona or play the most crucial role in the NP coating and uptake.

\subsection{Adsorption of proteins onto nanoparticles}
\label{sec::potentials}

The starting point for development of a CG model for the interaction of NPs with proteins is to decide how
much detail from the molecular structure of the protein one needs to keep. There an active and an extensive research
activity on the different CG models that can be used to simulate proteins under different conditions and
for detail reviews see~\cite{Noi2013,Toz2005,Tak2012}. The aim of this work is to propose a set of tools that could be used to
simulate the interaction of one or more proteins (and in some cases quite big proteins) with a NP, for relatively long
timescales. To meet this goals with a reasonable computational effort, the number of beads representing the protein should be
as small as possible but the proposed model should also preserve enough structural information about the molecule.
For these reasons we propose a single bead per residue model and consider the structure of the protein as a rigid body.
We have studied the predictions of this model in more detail in Ref. \cite{Lopez2015}.
The crystal structures of the proteins are obtained from the literature and one bead is per amino-acid is placed at the
position of the $\alpha$-carbon. At the end of this section we will test the validity of this first approximation.
The second approximation is what level of detail will be
needed to represent the NP. In this work, we will consider spherical homogeneous NPs so a single bead representation
is justified.

In our model, the total NP surface-protein potential interaction ($U$) is a function
of distance of the surface to the centre of mass (COM) of the protein, $d_{\mathrm{COM}}$ and of protein orientation.
It is given by a sum of two contributions:
\begin{equation}\label{eq::U}
U = \sum_i^{N} \left( U_i^{\mathrm{VdW}} + U_i^{\mathrm{el}}\right)
\end{equation}
where $N$ is the total number of residues in the protein, $U_i^{\mathrm{VdW}}$ is the van der Waals interaction
of residue $i$ with the surface and $U_i^{\mathrm{el}}$ is the electrostatics interaction of residue $i$ with the surface.

For van der Waals contribution to the potential energy we propose a modified version of the residue-residue interaction potential
as suggested in \cite{BerDes2009}. The model is based on the widely used residue-residue interaction energies proposed
by Miyazawa and Jernigan~\cite{MiyJer1996}, but instead of having a $20 \times 20$ interaction matrix
this is reduced to a table of normalized hydrophobicities, $\epsilon_i$, one for each amino acid
(see Table~II in \cite{BerDes2009}). A hydrophobicity index 0 is assigned to the most hydrophilic
residue (LYS) and an index 1 to the most hydrophobic one (LEU).
We should stress that any other hydrophobicity scale can also be used,
it just has to be transformed such that the indexes have to be
between 0 and 1, where 0 is assigned to the most hydrophilic residue while 1 to the most hydrophobic one.
In this work, we consider a general surface which chemical reactivity that can be
modeled  as another residue with a hydrophobicity index $\epsilon_s$.

To model interaction of biomolecules with particles of different sizes, we use the following model for the nanomaterial.
We assume that the interaction between a residue $i$ and a bead of the nanomaterial $s$ being at a distance $r$ from each other is given by a modified 12-6 Lennard-Jones potential:
\begin{equation}\label{eq::LJRS}
U_{s,i}(r) =
\left\{
        \begin{array}{ll}
                4\epsilon_{e,n}\left[\left(\frac{\sigma_{s,i}}{r}\right)^{12}-\left(\frac{\sigma_{s,i}}{r}\right)^{6}\right]
                +\epsilon_{e,n}(1-\epsilon_{s,i}) & r < r_{c,i}\\
               4\epsilon_{e,n}\epsilon_{s,i}\left[\left(\frac{\sigma_{s,i}}{r}\right)^{12}-\left(\frac{\sigma_{s,i}}{r}\right)^{6}\right] & r_{c,i}\leq r \leq r_{\mathrm{cut}} \\
                0 & r >  r_{\mathrm{cut}}
        \end{array}
\right.
\end{equation}
$\epsilon_{e,n}$ is a parameter that scales the interaction energy, $\epsilon_{s,i}$ is the combined hydrophobicity index of residue $i$ and the nanomaterial and is given by $\epsilon_{i,s}=\sqrt{\epsilon_i \epsilon_s}$, $\sigma_{s,i}$ is the average van der Waals radius of residue $i$ and the nanomaterial bead, $\sigma_{s,i}=(\sigma_s + \sigma_i)/2$, $r_{c,i}$ is the position of the minimum of the pair potential.

An integration of the 12-6 potential over the volume of the nanomaterial as defined in~\cite{BerDes2009} gives a 9-3 Lennard-Jones-type potential.
For a flat surface, the interaction can be expressed in terms of $d$, the distance between the residue centre of mass the closest element of the surface.
An integration over a semi-space gives:
\begin{equation}\label{eq::LJRS1}
U_{si}^{\mathrm{vdW}}(d) =
\left\{
        \begin{array}{ll}
                 \epsilon_{es} \rho \sigma_{s,i}^3 \left[\left(\frac{\sigma_{s,i}}{d}\right)^{9}-\frac{15}{2}\left(\frac{\sigma_{s,i}}{d}\right)^{3}
                +\left(\frac{125}{2}\right)^\frac{1}{2}(1-\epsilon_{s,i})\right ] & d < d_{c,i}\\
                 \epsilon_{es}  \epsilon_{s,i} \rho  \sigma_{s,i}^3 \left[\left(\frac{\sigma_{s,i}}{d}\right)^{9}-\frac{15}{2}\left(\frac{\sigma_{s,i}}{d}\right)^{3}\right] & d_{c,i}\leq d \leq d_{\mathrm{cut}}\\
                0 & d >  d_{\mathrm{cut}}
        \end{array}
\right.
\end{equation}
where $\epsilon_{es} = \frac{4 \pi }{45}\epsilon_{e,n}$, $\rho$ is the number density of beads in the nanomaterial, $d$ is the distance from the residue $i$ to the surface,
$d_{c,i} = (2/5)^{1/6}\sigma_{s,i}$. Although the density $\rho$ seems to be an important parameter scaling the interaction, it is not independent and therefore is not crucial for our method. From fitting the adsoprtion enrgy to experimental or MD simulation data, we can find the composite quantity $\epsilon_{es} \rho$, which is sufficient for further calculations.
For a nanoparticle of radius $R$, a similar integration over the particle volume gives:
\begin{equation}\label{eq::LJRS2}
U_{si}^{\mathrm{vdW}}(r) =
\left\{
        \begin{array}{ll}
                4 \epsilon_{es} \rho \sigma_{s,i}^3  \left[\frac{\left ( 15 r^6 R^3+ 63 r^4 R^5 + 45 r^2 R^7 + 5 R^9 \right) \sigma^{9}_{s,i}} {\left(r^2 - R^2\right)^9}
                - \frac{15 R^3 \sigma^{6}_{s,i}} {\left (r^2 - R^2 \right )^3} \right]
                - U^{\mathrm{vdW}}_c (1-\epsilon_{s,i}) & r < r_{c,i}\\
                4 \epsilon_{es}\epsilon_{s,i} \rho \sigma_{s,i}^3 \left[\frac{\left ( 15 r^6 R^3+ 63 r^4 R^5 + 45 r^2 R^7 + 5 R^9 \right) \sigma^{9}_{s,i}} {\left(r^2 - R^2\right)^9}
                - \frac{15 R^3 \sigma^{6}_{s,i}} {\left (r^2 - R^2 \right )^3} \right] & r_{c,i}\leq r \leq r_{\mathrm{cut}}\\
                0 & r >  r_{\mathrm{cut}}
        \end{array}
\right.
\end{equation}
where $r$ is the distance from residue $i$ to the centre of the nanoparticle. The distance $r_{c,i}$ corresponds to the minimum of the potential and $U^{\mathrm{vdW}}_c$ is the value of the function $U_{s,i}^{\mathrm{vdW}}(r_{c,i})$ as defined in the range $r_{c,i}\leq r \leq r_{\mathrm{cut}}$. We do not show the general expression for the position of the minimum as it is too bulky. The minimum is located at
$r_{c,i} - R \approx(2/5)^{1/6}\sigma_{s,i}$ at $R \gg \sigma_{s,i}$ and is displaced to shorter distances at smaller $R$. The variation, however, is not very large, at $R \to \infty$, $r_{c,i} - R \approx 0.858374 \sigma_{s,i}$, at
$R = 200 \sigma_{s,i}$ it is  $0.858375 \sigma_{s,i}$, at $R = 20 \sigma_{s,i}$ it is $0.858469 \sigma_{s,i}$, at $R = 2 \sigma_{s,i}$ it is $0.865242 \sigma_{s,i}$.

Note that the potential in Eq.~(\ref{eq::LJRS}) will only give a repulsive interaction between a highly hydrophilic surface and any residue (\textit{i.e.} defining $\epsilon_s=0$, gives $\epsilon_{s,i}=0$ for all residues).
On the other hand, assigning a non-zero value for $\epsilon_s$ will only change the magnitude of
the interaction between any residue and the surface but not the shape of the potential.
In this way, the proposed potential is limited to model only hydrophobic surfaces.
Because of this limitation, we set the value of $\epsilon_s=1$ for all simulations.
Alternatively, a potential that includes desolvatation
penalties, as the 12-10-6 Lennard-Jones potential proposed in~\cite{KTMH2008,KimHum2008} for residue-residue
interactions or the modified version proposed in~\cite{WeiKno2013} used to model residue-surface interactions, can be
used to generate a more general interaction potential.
The main drawback of the use of these more refine formulas for the potential is that the parameterisation
is more challenging, and the applicability of a set of parameters could be very narrow.

The electrostatic interactions in Eq.~(\ref{eq::U}) is modeled by adding point charges on the NP surface.
This charges interact with the charged residues via a Debye-H\"uckel potential.
The electrostatic interaction energy between a residue $i$ and all the charges on the surface is
given by:
\begin{equation}\label{eq::DHpotential}
U^{el}_{i} =  \sum_j^{N_e} \lambda_B k_B T q_i q_j \frac{\exp(-r_{ij}/\lambda_D)}{r_{ij}}
\end{equation}
where $r_{ij}$ is the distance between the residue $i$ and the point charge on the surface $j$,
$\lambda_B = e^2/ \left(4\pi\varepsilon_0 \varepsilon_r k_B T\right)$ is the Bjerrum length, $k_B$ is the Boltzmann constant,
$T$ the temperature, $\varepsilon_0$ the dielectric permittivity of vacuum, $\varepsilon_r$ the relative  dielectric permittivity of water,
$q_i$ the charge of residue $i$, $q_j$ the charge of the point charge $j$ on the surface, $N_e$ the total
number of point charges on the surface and $\lambda_D$ is the Debye length
(defined through $\lambda_D^{-2} = 8 \pi \eta \lambda_B c_0$, with $c_0$ is the background
electrolyte concentration). In practice, the points charges are evenly distributed on the spherical surface of the NP using
a Golden Section spiral algorithm and all points will have the same charge $q_j$
given by $q_j = 4 \pi \sigma R^2/N_e$, where $\sigma$ is the surface charge density of the NP
and $R$ is the radius of the NP.

\begin{table}
\begin{tabular}{c | c c c c c c c c c c}
\hline
\hline
Residue & LYS & GYU & ASP & ASN & SER & ARG & GLU & PRO & THR & GLY\\
$\epsilon_i$~$(\mathcal{E})$ & 0.00 & 0.05 & 0.06 & 0.10 & 0.11 & 0.13 & 0.13 & 0.14 & 0.16 & 0.17\\
$\sigma_i$~(nm) & 0.64 & 0.59 & 0.56 & 0.57 & 0.52 & 0.66 & 0.60 & 0.56 & 0.56 & 0.45\\
\hline
\hline
Residue & HIS & ALA & TYR & CYS & TRP & VAL & MET & ILE & PHE & LEU\\
$\epsilon_i$~$(\mathcal{E})$ & 0.25 & 0.26 & 0.49 & 0.54 & 0.64 & 0.65 & 0.67 & 0.84 & 0.97 & 1.00\\
$\sigma_i$~(nm) & 0.61 & 0.50 & 0.65 & 0.55 & 0.68 & 0.59 & 0.62 & 0.62 & 0.64 & 0.62\\
\hline
\hline
\end{tabular}\caption{Normalized hydrophobicities $\epsilon_i$
(taken from Table~II in \cite{BerDes2009} and $\sigma_i$ for each amino acid.
The most hydrophilic residue has a $\epsilon_i$ of 0, while the most hydrophopic
has a value of 1. For aminoacid-aminoacid interactions, we use the the Lorentz-Berthelot mixing
rules $\sigma_{i,j}=(\sigma_i + \sigma_j)/2$, $\epsilon_{i,j} = \sqrt{\epsilon_i \epsilon_j}$.}
\label{tab::parameters}
\end{table}

\subsection{Orientational sampling and the calculation of the adsorption energy}
\label{sec::Ead}

As we are assuming that the proteins are rigid bodies, we are not
considering conformational changes during the adsorption process.
Although, the adsorption process might conduce to conformational changes,
this events happen at longer times than orientational
changes on the surface~\cite{ARS2005}. Taking this into account, the
adsorption energies calculated here will give an valuable insight into the long
time formation of the NP-protein corona formation.

In our CG model, each residue of a protein is represented by a single bead located at the
$\alpha$-carbon position. The native structures are obtained from the Protein Data Bank,
and in Table~\ref{tab::PDB}
we report the proteins studied in this work, the PDB ID from which
the CG model were built and the abbreviation that will be used in the rest
of the text. The chosen proteins are some of the most abundant in human blood
and will have a major influence in the formation of the NP protein-corona.

\begin{figure}[tbh]
\centering
\includegraphics[width=\hsize]{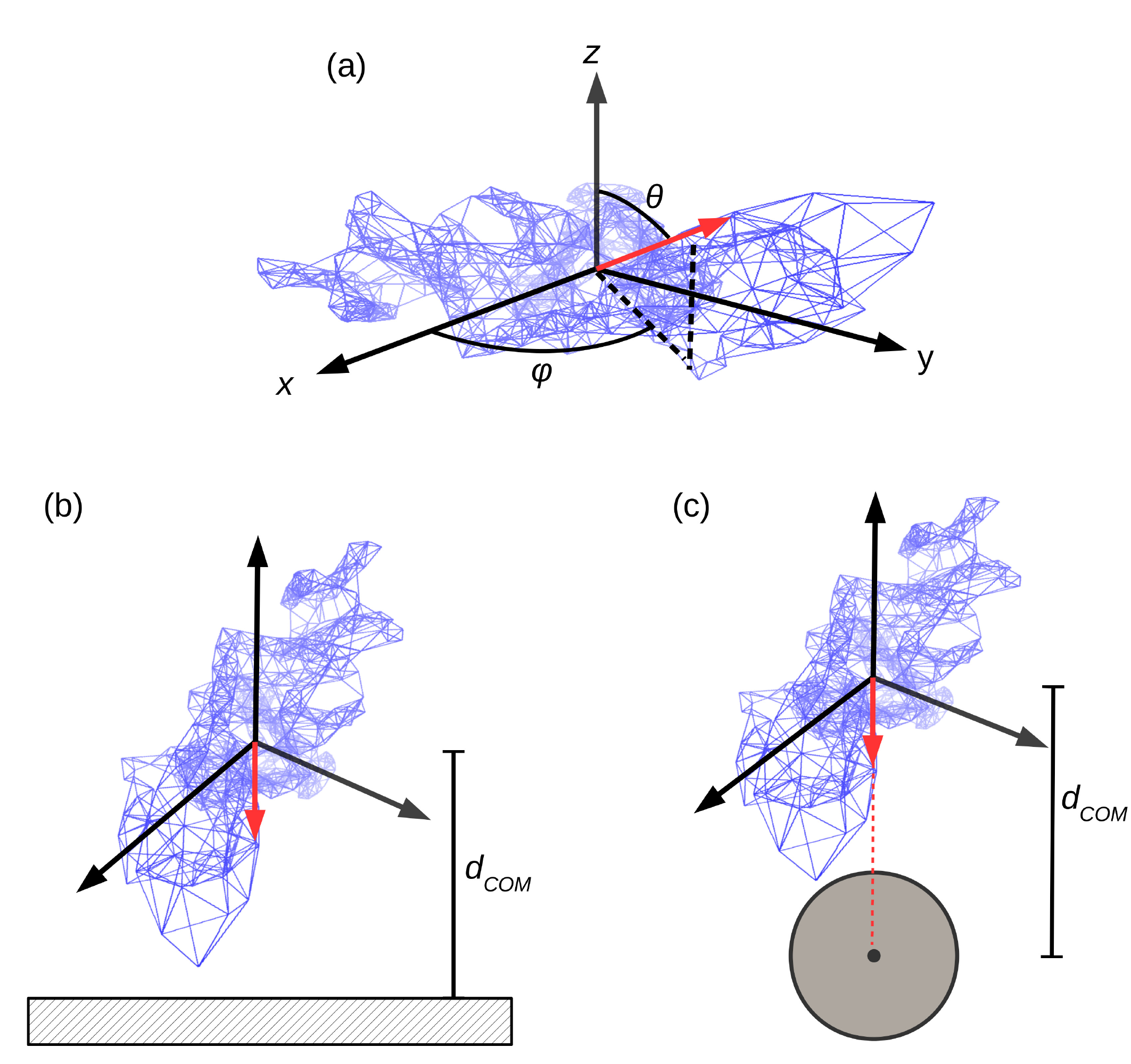}
\caption{Definition of the protein orientation. (a): Any point on the surface of the protein can be defined by a position vector
from the COM to that point and depends on two angles $\phi$ and $\theta$. The remaining degree of freedom is the
distance of the COM, $d_{\mathrm{COM}}$ to (b) the surface for a slab or (c) to the center of the NP for a NP.}
\label{fig::confispace}
\end{figure}

To identify the most favourable orientation of adsorbed protein globule
(the one with the minimum adsorption enthalpy) we will follow
the method suggested in~\cite{SWL2005}, which is not as efficient as \textit{e.g.}
a genetic algorithm, but can provide additional information about the adsorption process.
Briefly, a configuration space search is performed, where a systematic rotation
of the protein allows us to build an adsorption map.
There are three degrees of freedom (DOF) that
have to be scanned. Fig.~\ref{fig::confispace} shows that any point on the surface of
the protein can be defined by a position vector from the COM of
the protein. This vector is characterized by two angles: $\phi$ and $\theta$ and by
rotating the molecule an angle $-\phi$ around the $z$  direction and then by an angle $-\theta+180\,^{\circ}$
around the $y$ axes will make the position vector point towards the surface. The third DOF
is the distance from the COM to the closest point of the surface, $d_{\mathrm{COM}}$.
Here, we sample $\phi$ from 0 to $350^{\circ}$ in steps
of $10^{\circ}$ and $\theta$ from 0 to $170^{\circ}$ in steps of $10^{\circ}$
(note that $\phi=0^{\circ}$ is equivalent
to $\phi=360^{\circ}$, and that $\theta=0^{\circ}$ is equivalent to $\theta=180^{\circ}$).
Instead of obtaining the ``real'' adsorption free energy by calculating the
potential of mean force for all orientations, we only calculate the
potential energy $U$ (given by Eq.~(\ref{eq::U})),
which is the sum of all the interactions between the surface and the protein.
As the adsorption energies are expected to be at least 5 times $k_B T$ and as the proteins
are assumed to be rigid, neglecting thermal fluctuations is clearly justified.
For each configuration ($\phi_i$, $\theta_j$), the total potential energy is calculated as
a function of distance of the COM, $U(d_{\mathrm{COM}},\phi_i,\theta_j)$, to the surface for the case of a slab (Fig.~\ref{fig::confispace}b) or
to the center of the NP for the case of a NP (Fig.~\ref{fig::confispace}c). Following a similar
approach as in~\cite{Dariaetal2010}, and denoting the reaction coordinate $d_{\mathrm{COM}}=z$,
the adsorption energy for any particular configuration in the case of a protein adsorbing on a flat surface is given by:
\begin{equation}
E(\phi_i,\theta_j) = -k_B T \ln \left[ \frac{1}{a(\phi_i,\theta_j)} \int_0^{a(\phi_i,\theta_j)} \exp(-U(\phi_i,\theta_j,z)/k_B T) dz \right]
\end{equation}
where $a(\phi_i,\theta_j)$ is the maximum interaction distance from the COM of the protein to the surface for the given orientation.
For the case of a NP-protein interaction, the mean interaction energy for any particular orientation is given by:
\begin{equation}\label{eq::EsadNP}
\begin{array}{l}
E(\phi_i,\theta_j) = -k_B T
\times \ln \left[ \frac{3}{(R+a(\phi_i,\theta_j))^3-R^3} \int_R^{R+a(\phi_i,\theta_j)} z^2\exp\left(\frac{-U(z,\phi_i,\theta_j)}{k_B T}\right) dz \right]
\end{array}
\end{equation}
Then the total mean adsorption energy of the system for both cases (slab and NP), $E_{ad}$, can be estimated by averaging over all adsorbed
states with Boltzmann weighting~\cite{SWL2005}:
\begin{equation}\label{eq::Esad}
E_{\mathrm{ad}} = \frac{\sum\limits_{i} \sum\limits_{j} P_{ij} E(\phi_i,\theta_j)}{\sum\limits_{i} \sum\limits_{j} P_{ij}}
\end{equation}
where $P_{ij}=\sin(\theta_j)\exp[-E(\phi_i,\theta_j)/k_B T]$ is the Boltzmann weighting factor.

\subsection{Details of the simulations, parametrization and validation}
\label{sec::Para}

All simulation were performed using Espresso~\cite{Espresso} and
the cutoff for the interaction potential in Eq.~(\ref{eq::LJRS}) was set to $r_{\mathrm{cut}}=6$~nm.
For all calculation the simulation box was taken big enough to fit the NP and the protein.
The method described here only involves the calculation of the total energy of the system given by Eq.~\ref{eq::U},
therefore a coupling to a thermostat is not required. After the CG model were built from the PDB files, the obtained structures were shifted so the COM of the molecules was in the origin of the frame of reference and this structure was defined as the $(\phi=0^{\circ}$,$\theta=0^{\circ})$ orientation.
With this definition the
first residue in the sequence of each protein will have the following $(\phi,\theta)$ angles: ($21.4^{\circ}$, $85.2^{\circ}$)
for HSA and ($132.1^{\circ}$, $46.4^{\circ}$) for Fib.

The units of the simulations are: lengths ($\mathcal{L}$) in nm,
energy ($\mathcal{E}$) in $k_B T\approx 4.15 \times10^{-21}~\mathrm{J}$ taking a temperature
of $T=300~\mathrm{K}$, for the mass unit ($\mathcal{M}$) we selected the average
mass of the 20 residues (ca. $110~\mathrm{Da}$) hence in our
simulations all residues have a mass of 1.
The values of $\epsilon_i$ and $\sigma_i$ can be found in Table~\ref{tab::parameters} and
as mentioned in Sec.~\ref{sec::potentials} we will only consider hydrophobic NPs with
$\epsilon_{s}=1$ and $\sigma_s=0.35$~nm.

NPs with negative surface charges as well as neutral NPs were considered.
For the negative charged cases, a surface charge density of $-0.02$ C/m$^2$ was used.
As explained in Sec.~\ref{sec::potentials}, the charged surfaces are modelled by
individual point charges. The surface density of these charged beads ($\sigma_c=N_e/R^2$)
was set to 4~nm$^{-2}$ for all the simulations, which gives \textit{e.g.} a $N_e=100$ for a
NP of $R=5$~nm.
Then, we assumed that each bead carries a charge of $-0.39e$, where $e$ is the elementary charge.
As we are considering physiological conditions, we use $\lambda_B=0.73$~nm and $\lambda_D=1$~nm.
Residue charges at this condition are $+e$ for LYS and ARG, $-e$ for ASP and GLU, and
$+0.5e$ for HIS. The rest of the residues are neutral.

The only free parameters of the model are $\rho\epsilon_{es}$ in Eq.~(\ref{eq::LJRS}),
and the parameterisation was done by systematically changing its
value to match experimental data of adsorption of Lysozyme on
hydrophobic surfaces reported by Chen \textit{et. al.}~\cite{CHL2003}.
The native structure for our CG model of Lysozyme was obtained from the PDB ID:2LYZ.
With $\rho\epsilon_{es}=1.972k_B T/nm^3$ we obtain a value of $-7.6 k_B T$ for the
adsorption energy (very close
to the experimental reported value of $-7.9 k_B T$).

To validate the parameterisation, the adsorption energy of Myoglobin (PDB ID: 1MBN used for the CG model)
was calculated using the same value of $\rho\epsilon_{es}$ obtained from the parameterisation.
In this way, a value of $-6.1k_B T$ was found for the adsorption energy of Myoglobin.
This value is slightly lower that the experimental value of $-7.6k_B T$ also reported by Chen \textit{et. al.}~\cite{CHL2003} but
reproduces the trend that Myoglobin adsorbs slightly weaker than Lysozyme to a hydrophobic surface.

\subsection{Protein adsorption energies} \label{sec::results}

\begin{figure}[]
\centering
\includegraphics[width=\hsize]{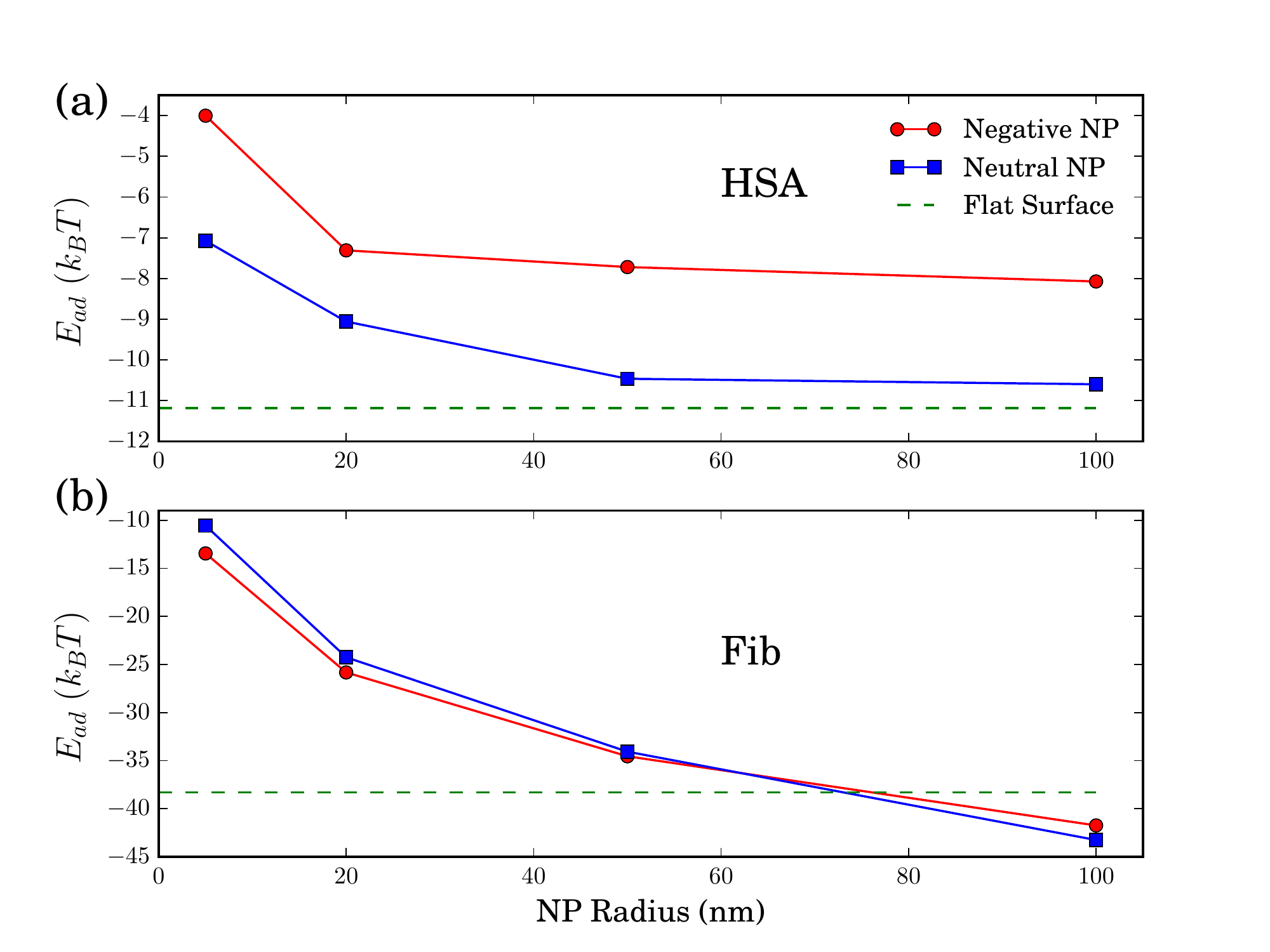}
\caption{Adsorption energies as a function of the NP radius for the proteins studied and the three types of surface charge:
Negative: $-0.02$ C/m$^2$ and Neutral: no charge. (a) HSA and (b) Fib. The dashed lines show the Adsorption energy
for the case of a flat surface.}
\label{fig::Eads}
\end{figure}

Results for the adsorption energies calculated using Eq.~(\ref{eq::Esad}) as a function of NP radius are shown in Fig.~\ref{fig::Eads}.
The results show that HSA adsorbs stronger as the radius of the NP increases until it reaches a minimum value (Fig.~\ref{fig::Eads}a).
For small NPs, the combination of the curvature effect (increasing $R$ increases the van der Waals interactions interactions)
with the availability of residues to interact with the surface originates that the proteins adsorbs stronger
(more negative values) as the radius is increased. Then after a value of radius around 50 nm the
$E_{\mathrm{ad}}$ starts to converge to the value corresponding to a flat surface as the van der Waals interactions interactions and
the number of residues close to surfaces do not change significantly by increasing $R$. We performed calculations for NPs of $R$ up to
500 nm and confirmed that the adsorption energy indeed converges to the slab values.
For the Fib molecule the situation is different (Fig.~\ref{fig::Eads}b). In this case the adsorption energy decreases as a function
of $R$ at least until the biggest radius studied here ($R=100$~nm) and it is lower that for the adsorption onto a flat surface.
The big size of the Fib molecule (ca. $45$~nm on its longest axes) makes that for at least until $R=100$~nm
the combined effects of curvature and number of residues that interact with the surface are still noticeable.
The effect of the charge is more important for the HSA that for the Fib. HSA is overall negative, so the electrostatic interactions
contribution is mainly repulsive increasing the values of the $E_{ad}$. On the other hand, the Fib molecule is positive and
the electrostatics interactions tend to increase the adsorption of the Fib onto a negative surface. In neither of the proteins
the effect of electrostatic interactions was more that than $3~k_B T$

\begin{figure}[]
\centering
\includegraphics[width= \hsize]{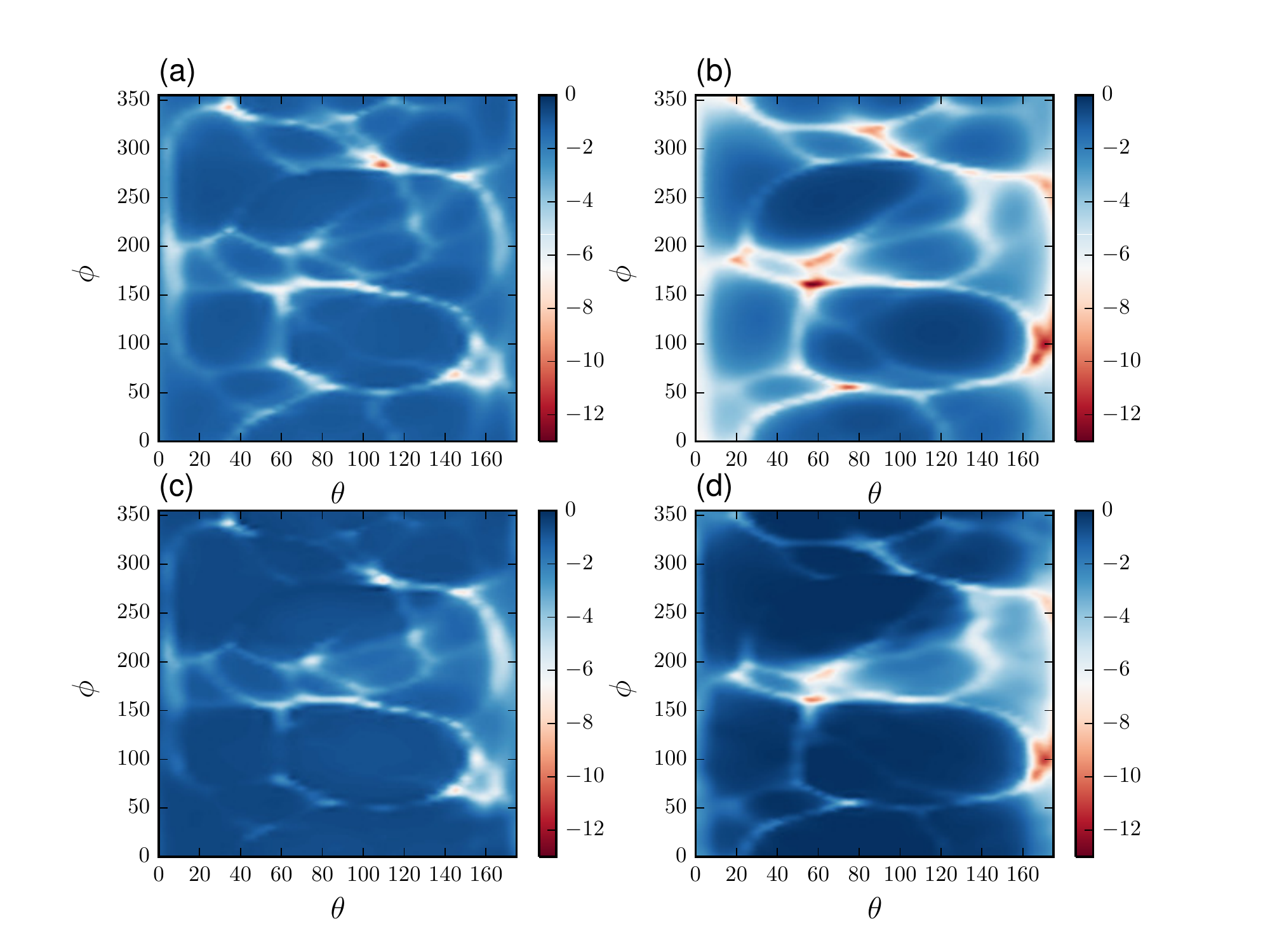}
\caption{Adsorption energy maps for HSA. (a) Neutral particle of $R=5$~nm. (b) Neutral particle of $R=100$~nm.
(c) Negatively charged particle of $R=5$~nm. (d) Negatively charged particle of $R=100$~nm.}
\label{fig::SMHSA}
\end{figure}

The systematic sampling employed for the calculation of the adsorption energies
can also be used to identify the most favourable orientations for adsorption and to study
how the charge and/or the radius of the NP influence the protein orientation.
Fig~\ref{fig::SMHSA}, shows a surface map of the adsorption energy as a function of the angles $\theta$ and $\phi$ for HSA.
Each panel is for a radius of 5 or 100~nm and for a neutral or a negative charged surfaces.
In general, the surfaces are complex in structure showing an energy landscape with several
local minima with differences less than $1k_B T$. It is also important to
notice that the maps have large areas with adsorption energies of $-6 k_B T$ or lower.
Our results show that HSA will strongly adsorb at physiological conditions and room temperature and that
orientational changes after adsorption are energetically favourable.
Comparison of different panels in Fig~\ref{fig::SMHSA} shows that radius has only a small effect on the preferred orientations,
while the charge density has an unnoticeable impact on the preferred orientations.

\begin{figure}[]
\centering
\includegraphics[width=\hsize]{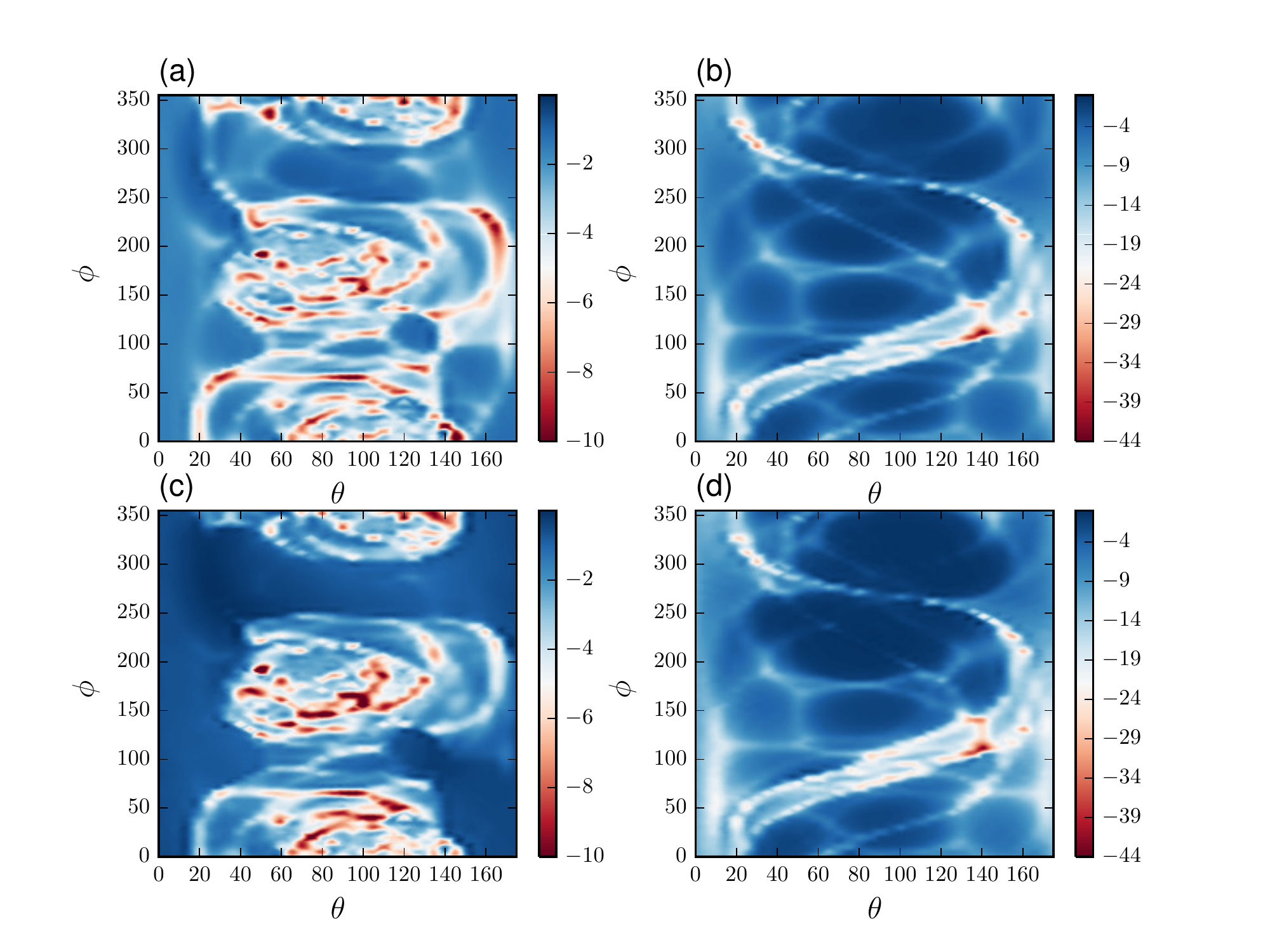}
\caption{Adsorption energy maps for Fib. (a) Neutral surface and $R=5$~nm.
(b) Neutral surface and $R=100$~nm. (c) Negatively charged surface and $R=5$~nm.
(d) Positively charged surface and $R=100$~nm.}
\label{fig::SMFib}
\end{figure}

\begin{figure}[]
\centering
\includegraphics[width=\hsize]{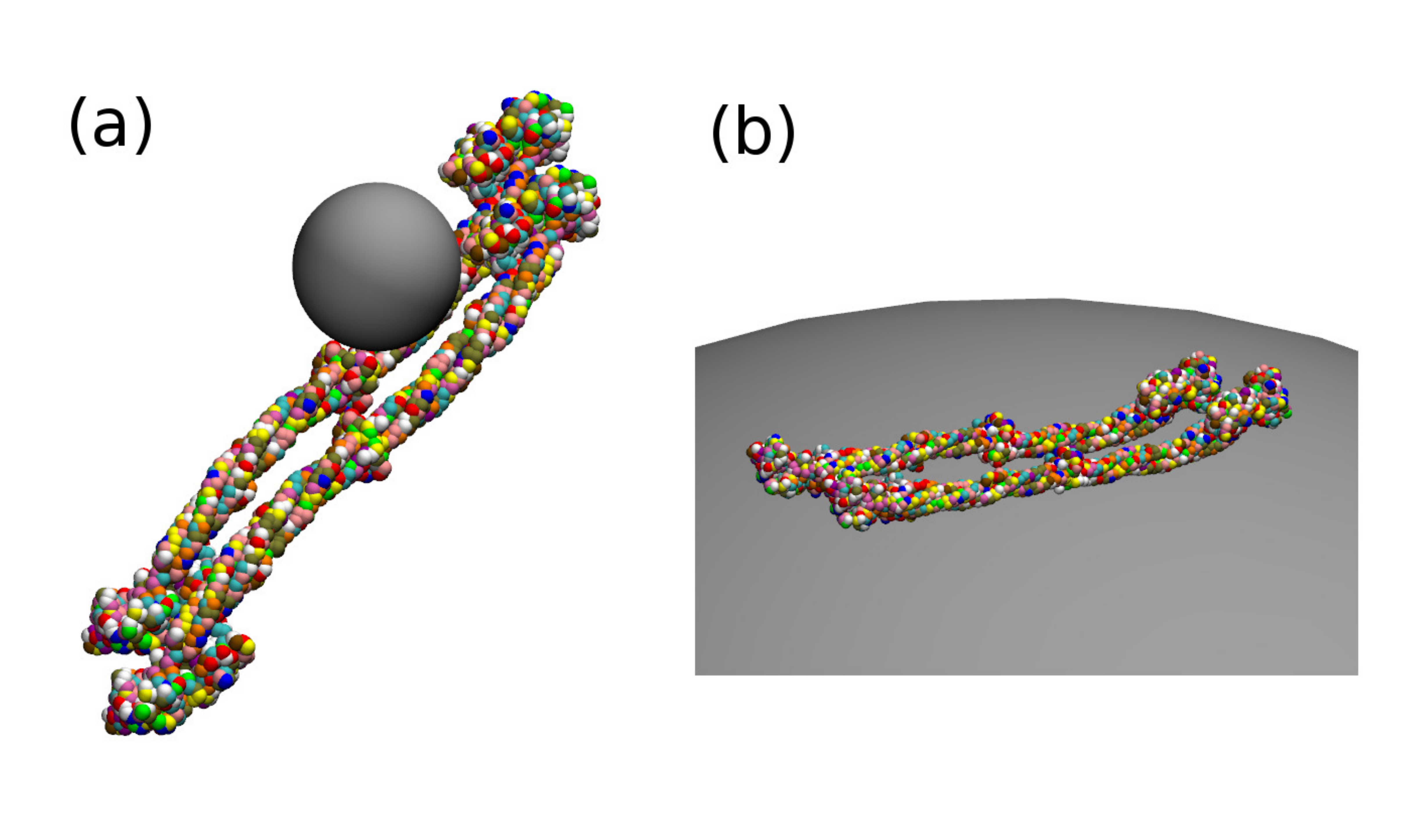}
\caption{Fib most favorable orientation for adsorption on a neutral surface: (a) $R=5$~nm and (b) $R=100$~nm.}
\label{fig::Fibscene}
\end{figure}

A different scenario is observed for Fib. Fig~\ref{fig::SMFib} shows
a surface maps of Fib adsorption for two radii for neutral and charged surfaces. In this case, the surface
maps depend on the radius of the NP (compare Fig~\ref{fig::SMFib}a with Fig~\ref{fig::SMFib}b or Fig~\ref{fig::SMFib}c with Fig~\ref{fig::SMFib}d)
but change very little between the charged and uncharged surface (compare Fig~\ref{fig::SMFib}a with Fig~\ref{fig::SMFib}c or Fig~\ref{fig::SMFib}b
with Fig~\ref{fig::SMFib}d). As we already noticed for HSA, the charge has a small effect on the total adsorption
energy so it is expected that it would not dramatically change the surface maps. The radius and the surface curvature seem to be more important
as for big proteins (like Fib) a larger NP allows a more extensive contact and thus influences the preference for protein orientations (or NP binding pockets).
In Fig.~\ref{fig::Fibscene} we show the most favorable orientations for Fib on a neutral
surface for two different NP radii. For the small NP (Fig.~\ref{fig::Fibscene}a), Fib has its
adsorption energy minimum in a configuration where the NP interacts with a relatively small segment of the molecule.
Meanwhile for a large NP, Fib tends to adsorb in a completely different orientation (Fig.~\ref{fig::Fibscene}b).
Now the most favourable orientation is one that oriented with the longest axis of the Fib molecule along the surface.

A very straightforward conclusion from the above data is that the bigger the protein, the stronger it will
adsorb onto a NP. This result agrees with the experimental observation reported by De Paoli~\textit{et. al}~\cite{Petal2010}, which
show that the binding association constant on citrate coated gold
NPs (which can be considered as moderately negative hydrophobic NP) depend mainly on the size of the protein (they studied HSA, Fib and other blood proteins).
It is interesting also to compare our results with the simulations of NP corona formation reported by Vilaseca \textit{et al.}~\cite{VDF2013}.
Using CG MD simulations they found that for a flat surface at long times the most abundant
protein adsorbed were Fib, then immunoglobulin-$\gamma$ (of intermediate size between HSA and Fib) and at last HSA.
At this point, it is important to remark that the adsorption energies calculated in this work
could be a good predictor of the equilibrium composition of the NP-protein corona
but at short times other factors such as the protein sizes and their concentrations
have to be considered to predict the corona composition.

\subsection{Validation of the methodology}
\label{sec::validation}

We now test our CG methodology with predictions of full-atomistic MD simulation. We model adsorption of small plasma protein
Ubiquitin (Ubi) to a flat TiO$_2$ surface. The reasons of choosing Ubi for the validation was due to the it size (only 76 residues)
and known folded structure, which allows us to perform full atomistic simulations in a reasonable amount of time.

\begin{figure}[]
  \includegraphics[width=0.3\textwidth]{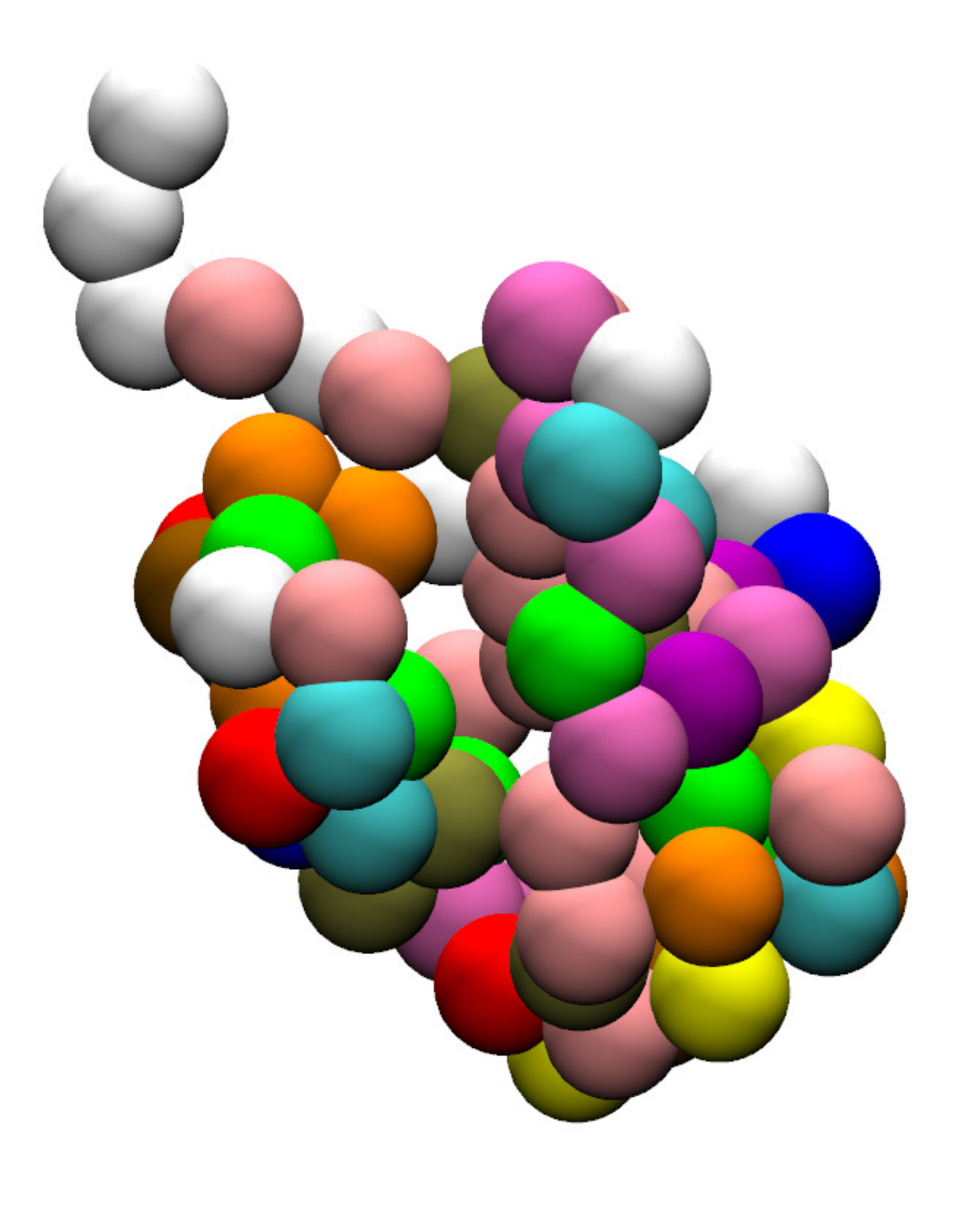}
  \caption{CG model of Ubi (PDB ID: 1Ubi~\cite{Vijay-Kumar1987}).
In our model each residue in the protein is represented by one bead}\label{fig::ubi}
\end{figure}

The Ubi crystal structure was obtained from the PDD (PDB ID file 1Ubi~\cite{Vijay-Kumar1987}) and was CG as explain
above (see Fig.~\ref{fig::ubi}). To be able to directly compare against full atomistic simulations, in this
case the potential interaction between the 20 different residues
and the surface were obtain by performing full atomistic simulations of the adsorption of each of the
20 aminoacids and then performing an inverse Monte Carlo.

\begin{figure}[]
  \centering
  \includegraphics[width=0.5 \textwidth]{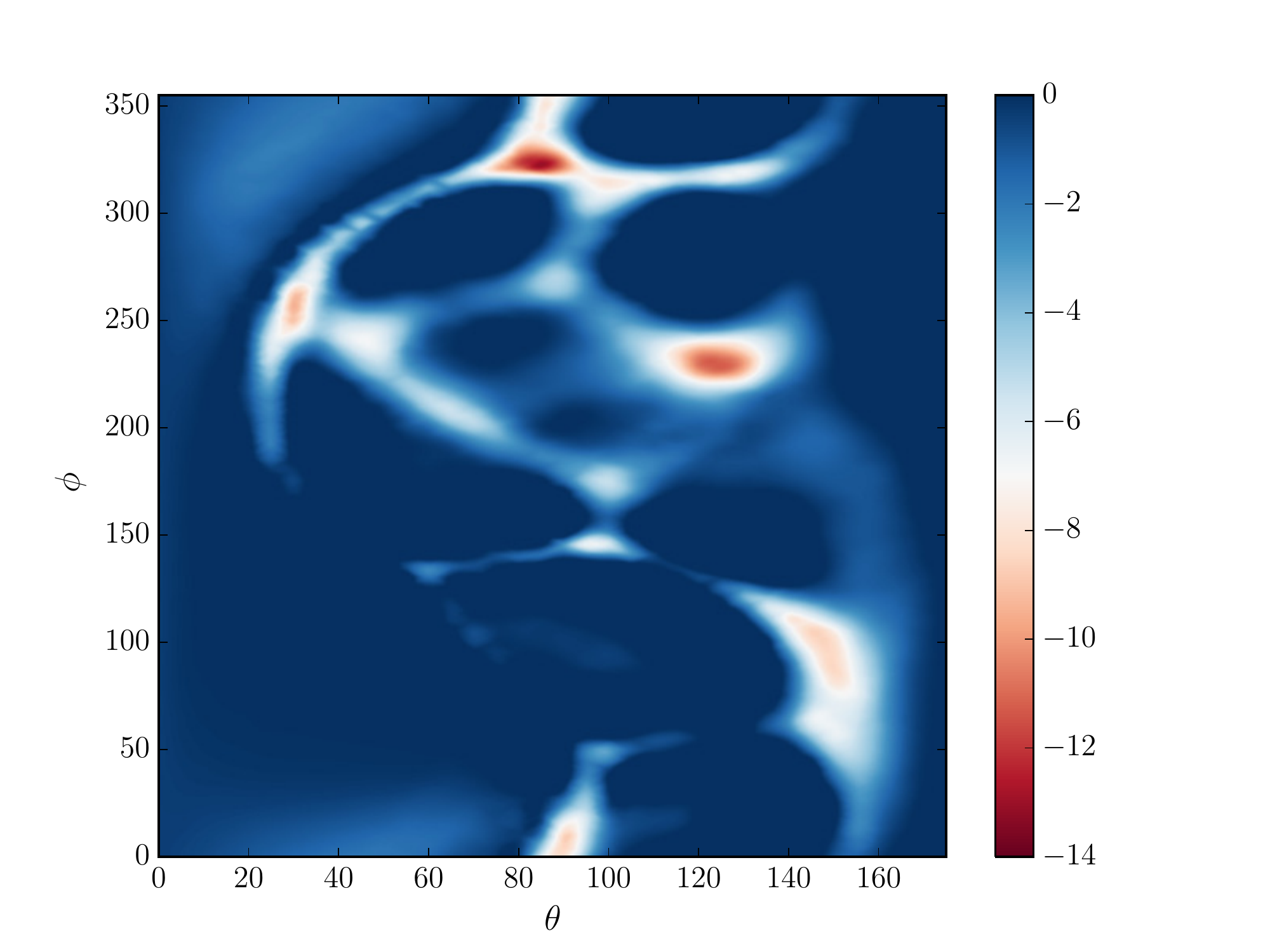}
  \caption{Adsorption energy maps for Ubi adsorbing into a TiO$_2$ slab.}\label{fig::mapUbi}
\end{figure}

With the CG methodology we obtained the total adsorption energy of $-10.7 k_BT$ for Ubi and the adsorption map is shown in Fig.~\ref{fig::mapUbi}.
The surface obtained shows two major minima and a number of local minima. It also predicts that orientational
changes are favorable after adsorption as some of the minima are connected by an energy landscapes with rather small
barriers (less than $3 k_BT$).

To analysis the validity of our model and to understand the dynamical behavior
of the adsorption process we now preform a series full atomistic simulations. We used
the VESTA program~\cite{Momma2011} to construct a $5 \times 20 \times
32$-supercell of the TiO$_2$ rutile unit cell. The coordinates were rotated so
that the normal of the TiO$_2$ slab, corresponding to the (100) surface, was
oriented along the $z$-direction. The box was elongated in the $z$-direction
and periodicities were kept in all directions. The final size of the simulation
box was then $9.466 \times 9.184 \times 12$ nm. Covalent bonds were added to
all Ti-O pairs within a 2 \AA-cutoff. We used force field parameters for the
TiO$_2$ slab from a recent parameterisation study~\cite{Brandt2014}.
This same force field was use to calculate the CG potential interactions between the
surface and the 20 amino acids.
The same folded Ubi structure as for the CG model was used and inserted above the TiO$_2$ slab.
The TiO$_2$-Ubi system was solvated by insertion of 25817 water molecules around
the protein and the slab, and the final system contained 99802 atoms. The
system was energy minimized for 1000 steps and then equilibrated at constant
temperature (300 K) and pressure (1 bar) for 100 ps using Berendsen's weak
scaling algorithms~\cite{Berendsen1984}, with relaxation constant $\tau=1$ ps in both
cases. The temperature coupling was applied independently to the TiO$_2$ slab
and to the rest of the system. The pressure tensor must be kept anisotropic due
to the solid TiO$_2$ slab, but the off-diagonal components of the
compressibility tensor (and the reference pressure tensor) were set to zero to
enforce a rectangular simulation box. The diagonal elements of the
compressibility tensor were set to $5 \times 10^{-7}$ bar$^{-1}$ in the lateral
directions (bulk TiO$_2$) and $5 \times 10^{-5}$ bar$^{-1}$ in the normal
direction (bulk water). The box vectors relaxed 2-4\% during equilibration.

In a first simulation, we placed the protein in the ($\phi=0^{\circ}$, $\theta=0^{\circ}$)
orientation close to the surface
and followed the dynamics for 440 ns at constant volume and
300 K. The Nose-Hoover thermostat~\cite{Hoover1985,Nose1984} with the coupling
constant $\tau=5$ ps was used to ensure proper sampling of the ensemble when
controlling the temperature. The simulation was run in parallel using 512 cores
and frames were kept every 5 ps.

\begin{figure}[t]
  \centering
  \includegraphics[width=0.5\textwidth]{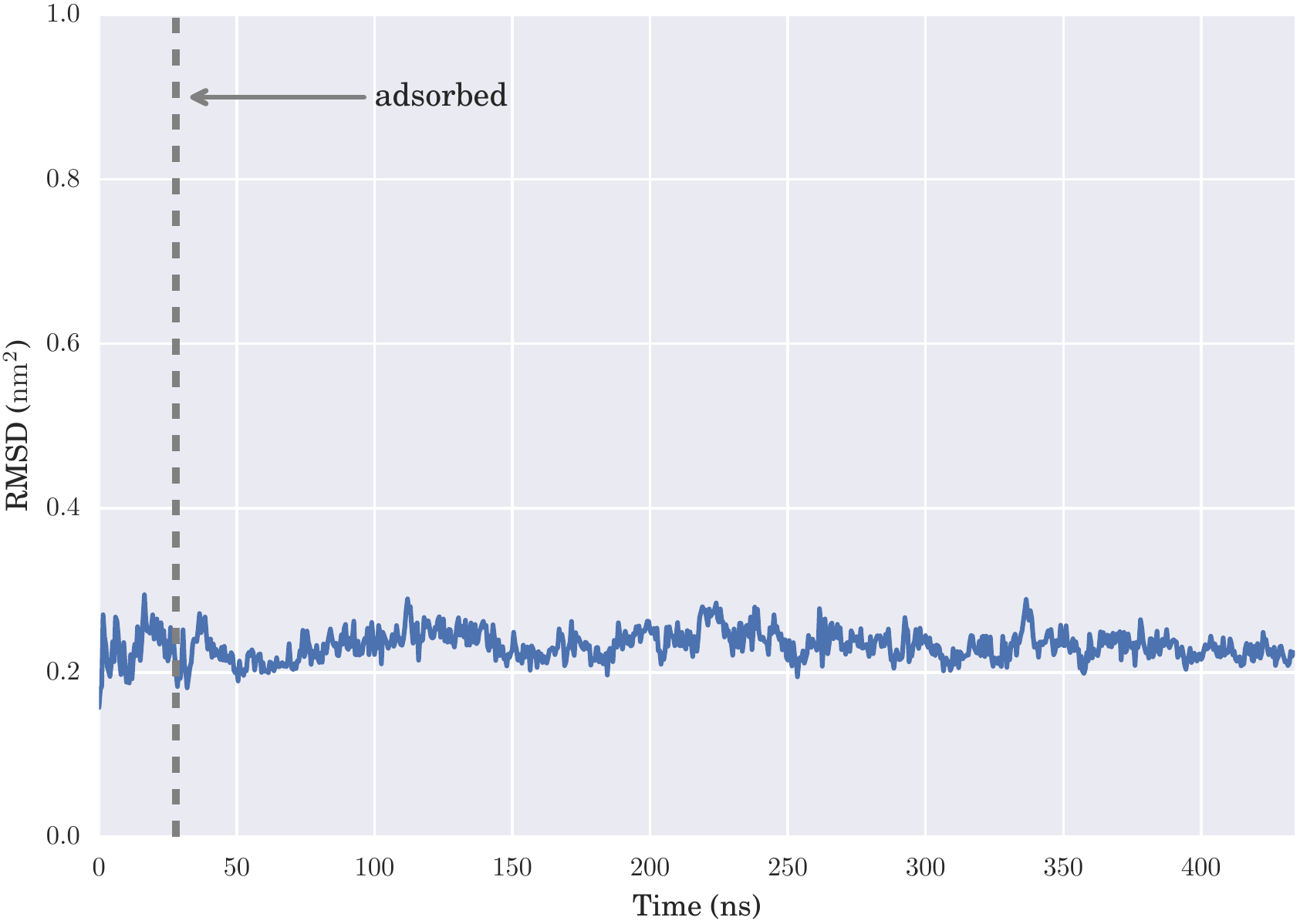}
  \caption{\label{fig:RMSD}The root-mean-square deviation (RMSD) of Ubi
    during the simulation with respect to the PDB reference structure. No
    unfolding occurs and the RMSD is $0.15$ nm$^2$ throughout the
    simulation, which is the same as found in simulations of the folded
    structure in bulk water.}
\end{figure}

The trajectory obtained showed that the protein motion was diffusive in the bulk
water for about 25 ns until making contact with the TiO$_2$
slab. Then the Ubi molecule attached to the surface and remained adsorbed for the rest of the
simulation. To study the stability of the structure of the
protein during adsorption we calculate the root-mean-square-deviation (RMSD) as a function of time and the results are shown
in Fig.~\ref{fig:RMSD}. The RMSD remained at a low constant
value of ca. 0.2 nm$^2$ during the simulation, i.e., no unfolding occurred in
the adsorbed state. This results clearly verifies that for the adsorption of Ubi on
TiO$_2$, a rigid body model for the protein structure is well justified.

\begin{figure}[t]
  \centering
  \includegraphics[width=0.9\textwidth]{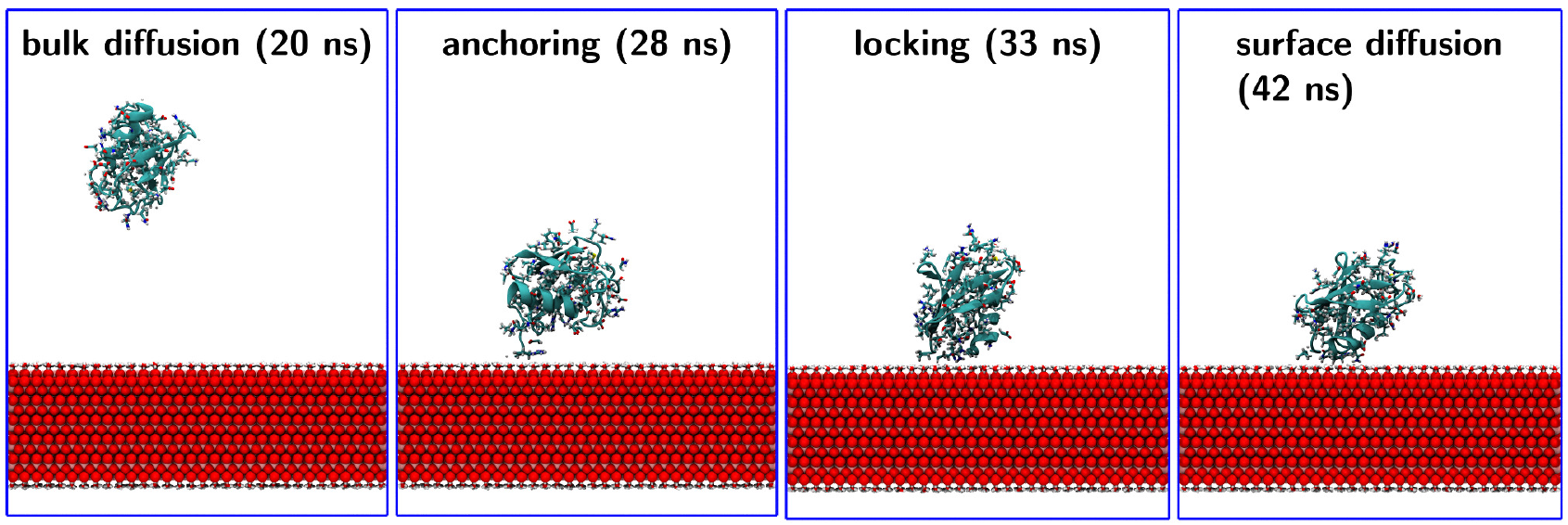}
  \caption{\label{fig:adsorption}Four distinct stages were identified during
    adsorption. (a) The protein diffuses in the bulk water. (b) The N-terminal
    of Ubi anchors to the solvation layer of the TiO$_2$ slab. (c) The protein
    rotates and locks to the solvation layer through GLN40 and GLN31. (d) The
    protein diffuses on the solvation layer in the locked conformation.}
\end{figure}

A detailed study of the simulation trajectory revealed that
the protein motion could be characterized by four states, and that adsorption
occurs through a two-step mechanism (Fig.~\ref{fig:adsorption}). First, the
protein diffuses freely in the bulk water. Second, the C-terminus of Ubi the
protein contacts the TiO$_2$ surface and provides a lock for the protein to the
first solvation layer. Third, Ubi rotates and locks into position on the
surface. Fourth, the protein diffuses on the surface in the locked orientation.

The adsorption mechanism is relatively fast once the first surface contact is
initiated. The anchoring of the C-terminus to the first solvation layer occurs
in about 5 ns and the locking is completed after 10 additional ns. The
residues at the C-terminus are ARG 74, GLY 75 and GLY 76. The
anchoring is initiated when the charged end of ARG74 contacts the surface
perpendicularly. The contact is not with the bare surface, but with the first
solvation layer, which is strongly bound directly to the surface. A this point
the rest of the protein diffuses in the bulk until (ca. 5 ns) GLU 40
can contact the surface, which leads to the protein being locked into an
adsorbed orientation after 10 ns. The locking procedure consists of Ubi
first connecting GLN40 to the surface, followed by a second connection through
GLN31. The two glutamines (GLN31 and GLN40) form a bridge that stabilize the
orientation of the protein and no more change in orientation occur for the rest
of the simulation.

The residues involved in the anchor-lock mechanism are arginine and
glutamine. These have been identified by potential of mean force
calculations~\cite{Brandt2014} of isolated side chain fragments (together with
aromatic side chains) to be the strongest binders to TiO$_2$. In both cases,
the NH$_2$-group of the end of the amino acid approaches the surface in a
perpendicular orientation, but can then rotate to maximize the interactions
with the solvation layer.

The ``upright'' position of the protein in the adsorbed state suggests that it
does not correspond to a free energy minimum. Since the orientation does not
change over 400 ns, it is likely that there are high free energy barriers
associated with the orientation changing into another free energy basin. To map
all values of $\phi$ and $\theta$, more sampling of the protein adsorption is
needed. This could either be done in a repetitive fashion from different
starting configurations or with enhanced sampling techniques such as
metadynamics.

As for Ubiquitin we do not observe any unfolding within several hundreds of nanoseconds,
we can conclude that our rigid protein model for studying adsorption is justified at least
for some conditions: small nanoparticles, non-metallic particles and small an compact proteins
such that the adsorption energies are within few tens of $k_BT$.
For other situations, one should evaluate the energy to decide whether the model is sufficient.
In general, conformational changes can be an important factor for the adsorption dynamical process~\cite{Petal2010}.
This assumption can be relaxed by \textit{e.g.} using a G\={o}--Type model (see~\cite{Noi2013} for a review on CG models
of proteins). Furthermore, as the methodology we presented is computationally very efficient
and can provide information about the structure of NP-protein complexes, it can be used as an exploring tool to
perform more sophisticated and computationally demanding calculations.

\section{Coarse-grained model of a lipid bilayer} \label{sec::bilayer}

Any attempt to simulate with some molecular details but at length and time scale involve in the
uptake of NPs through a cell membrane must rely on a CG model of the main
constituents of the biological membranes. In this section we describe a methodology to systematically CG
a lipid bilayer and lipid bilayer containing cholesterol from the results of full atomistic simulations.

\subsection{Molecular simulations of various lipid-cholesterol mixtures}\label{sec::I}

We started the CG procedure by performing all-atom molecular dynamics simulations for three lipid mixtures:
(i) 1,2-dimyristoyl-sn-glycero-3-phosphatidylcholine with cholesterol (DMPC + CHOL);
(ii) 1,2-dioleoyl-sn-glycero-3-phosphatidylserine with cholesterol (DOPS + CHOL);
(iii) 1,2-dimyristoyl-sn-glycero-3-phosphatidylcholine with 1,2-dioleoyl-sn-glycero-3-phosphatidylserine
(DMPC + DOPS). The composition of this systems are reported in the Table~\ref{tab:sys}.
In each simulation, the starting state was generated randomly and was energy
minimized afterwards. Then a short 1 ns NVT simulation at density 1 g/cm$^3$ have been carried out, which was
followed by 100 ns equilibration simulation in NPT-ensemble and production stage of 400 ns.
The Slipids force field was used~\cite{JoaLyu2012,JoaLyu2013}. Other simulation parameters: time step 2 fs;
Nose-Hoover isotropic thermo/barostat with temperature 303 K, pressure 1 bar,  relaxation
times 0.1 and 1 ps for thermostat and barostat respectively; all bonds were constrained by
Links algorithm; particle-mesh Ewald with Fourier spacing 1 \AA~ and tolerance parameter
$10^{-5}$. The configurations were saved in the trajectory each 10 ps. The atomistic
simulations were performed using the Gromacs simulation engine (v. 4.5) and a rigid TIP3P water model.

\begin{table}
\begin{tabular}{ c c c c}
\hline
\hline
System & I. DMPC-cholesterol & II. DMPC-DOPS & III. DOPS-cholesterol \\
\hline
Number of DMPC & 30 & 30 & - \\
Number of DPPS & - & 30 & 30 \\
Number of cholesterol & 30 & - & 30 \\
Number of water & 2000 & 2000 & 2000 \\
Number of Na+ & - & 30 & 30 \\
\hline
\hline
\end{tabular}\caption{Composition of the simulations used for the CG of lipids mixtures.}
\end{table}\label{tab:sys}

\subsection{Mapping of atomistic to  coarse-grained trajectories -- from residue to beads}\label{sec::II}

\begin{figure}[]
\centering
\includegraphics[width=\hsize]{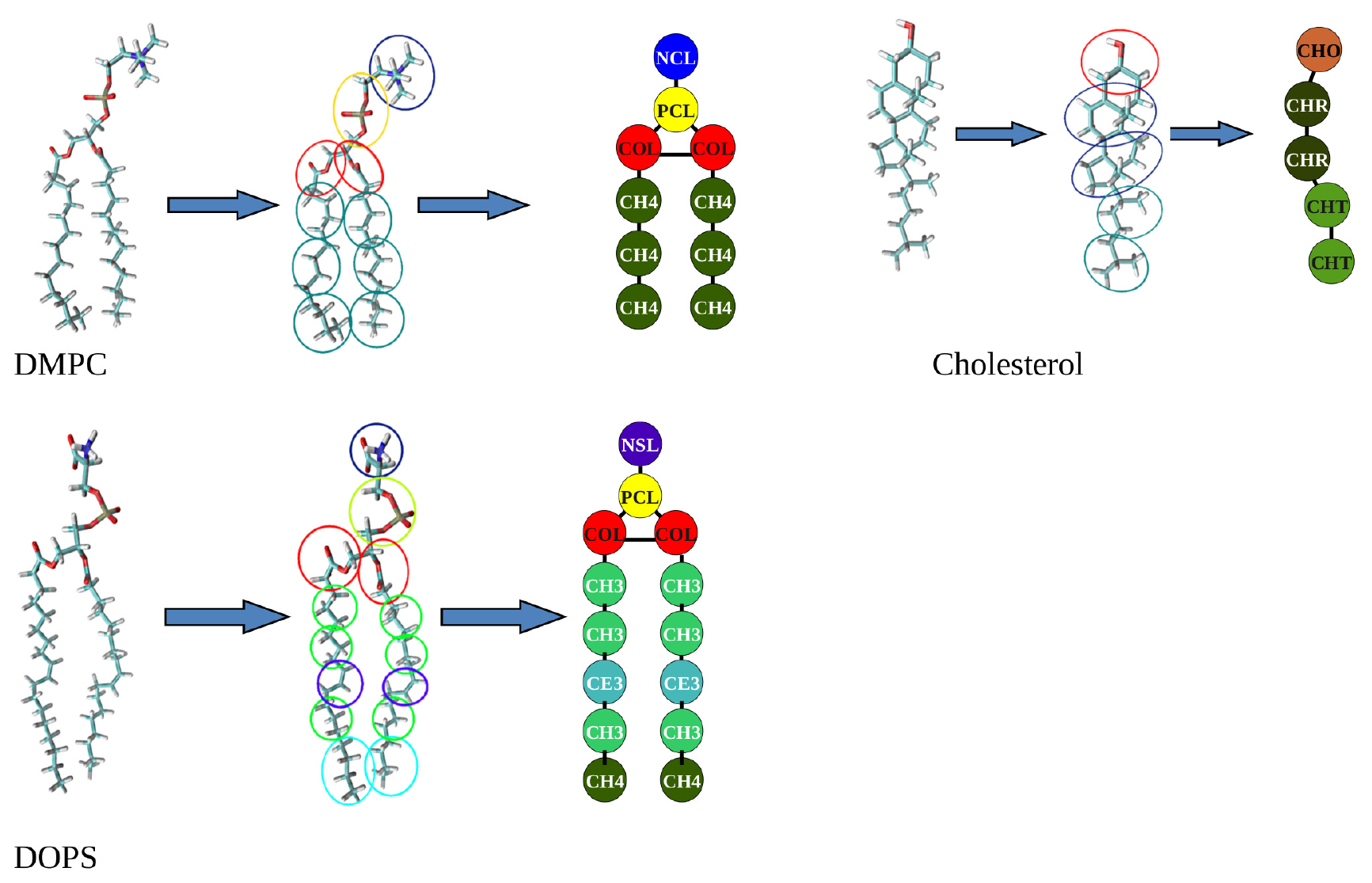}
\caption{Mapping of systems at an atomistic level to a coarse grained level where each
residue of the atomistic system is replaced by a bead for DMPC,
Cholesterol and DOPS (1,2-dioleoyl-sn-glycero-3-phosphatidylserin)
molecules. Coarse-grained sites of the same type are given by the same color.}
\label{fig::figD4.2}
\end{figure}

\begin{figure}[]
\centering
\includegraphics[width=\hsize]{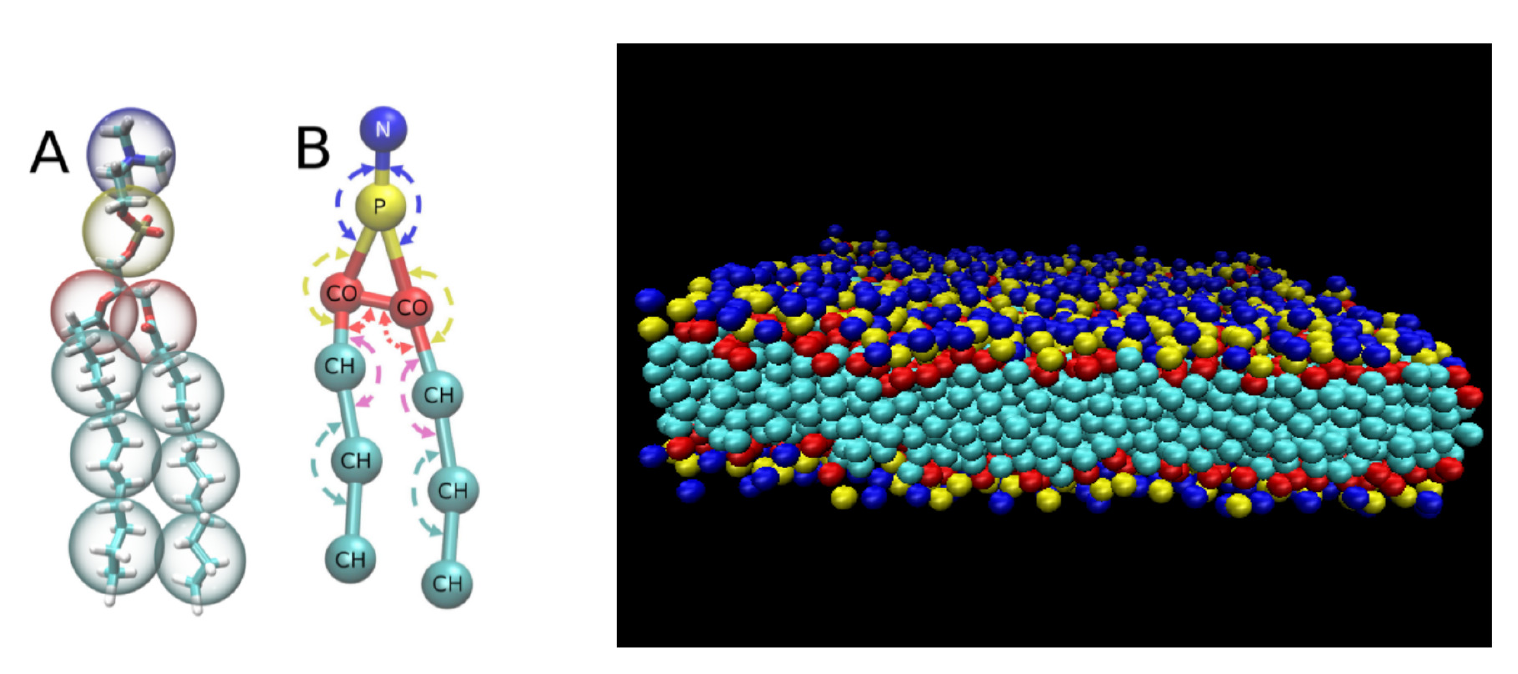}
\caption{Simulation snapshot of a single coarse-grained DMPC lipid molecule (left) and of self-assembled
DMPC bilayer 15x15 nm containing 762 lipids (right).}
\label{fig::figD4.3}
\end{figure}

The atomistic trajectories obtained in the simulations were mapped onto coarse-grained trajectories,
and radial distribution functions between sites of the coarse grained models have been determined.
As shown in Fig.~\ref{fig::figD4.2}, 10 beads for representation of DMPC molecule were used at the CG level (3 beads
instead of each of the two hydrocarbon tails,  4 beads instead of the head group including esters),
14 beads for DOPS molecule (5 beads instead of each hydrocarbon tail with specific distinguishing of
the beads with double bond and beads uniting 3 or 4 methylene groups, and 4 beads instead of the head group),
5 beads for CHOL molecule, and Na+ ions as a single bead were used. Water was not included into
CG model but its effect was included into solvent-mediated potentials. Fig.~\ref{fig::figD4.3} shows
a snapshot of  CG DMPC bilayer, which is spontaneously formed in a CG lipid system.


The radial distribution functions (RDF) between CG sites obtained after coarse-graining of the
atomistic trajectories were used to compute effective potentials defining interactions in the
coarse-grained models using the inverse Monte Carlo method. The RDF were computed for each pair
of different coarse-grained sites and were used as an input to compute effective coarse-grained
potentials  which reproduce the RDFs. Computations of effective potentials were done for the
same compositions of the systems I, II, and III as the respective atomistic simulations listed in Table~\ref{tab:sys}.
The software package MagiC~\cite{MirLyu2013} was used. In the inversion process, the first 20 iterations
have been carried out using iterative Boltzmann inversion, followed by 30-40 iterations using the
inverse Monte Carlo algorithm.

More specifically,  the RDF's have been determined between beads involved in ``non-bonded'' interactions,
that is between CG sites belonging different molecules or the same molecule but separated by more than two bonds.
Also, reference distribution functions for the bond lengths and bending angle distribution functions
were determined for the all CG sites relevant for the three types of considered molecules.
Then the calculated RDFs and bonded reference distribution functions were used to calculate
parameters of the corresponding coarse-grained potentials. This is a multistage process,
from a high resolution system description to a low resolution one.
Monte Carlo computer simulations of the coarse-grained system DMPC + CHOL using Metropolis method
(MagiC package) were carried out. The parameters were calculated using a two step iteration technique:
first, the iterative Boltzmann inversion method was performed to calculate a set of intermediate parameters;
second, the inverse Monte Carlo algorithm was used to calculate the final set of parameters.
The final parameters of the coarse-grained potentials for DMPC + CHOL mixture have been calculated.

\begin{figure}[]
\centering
\includegraphics[width=\hsize]{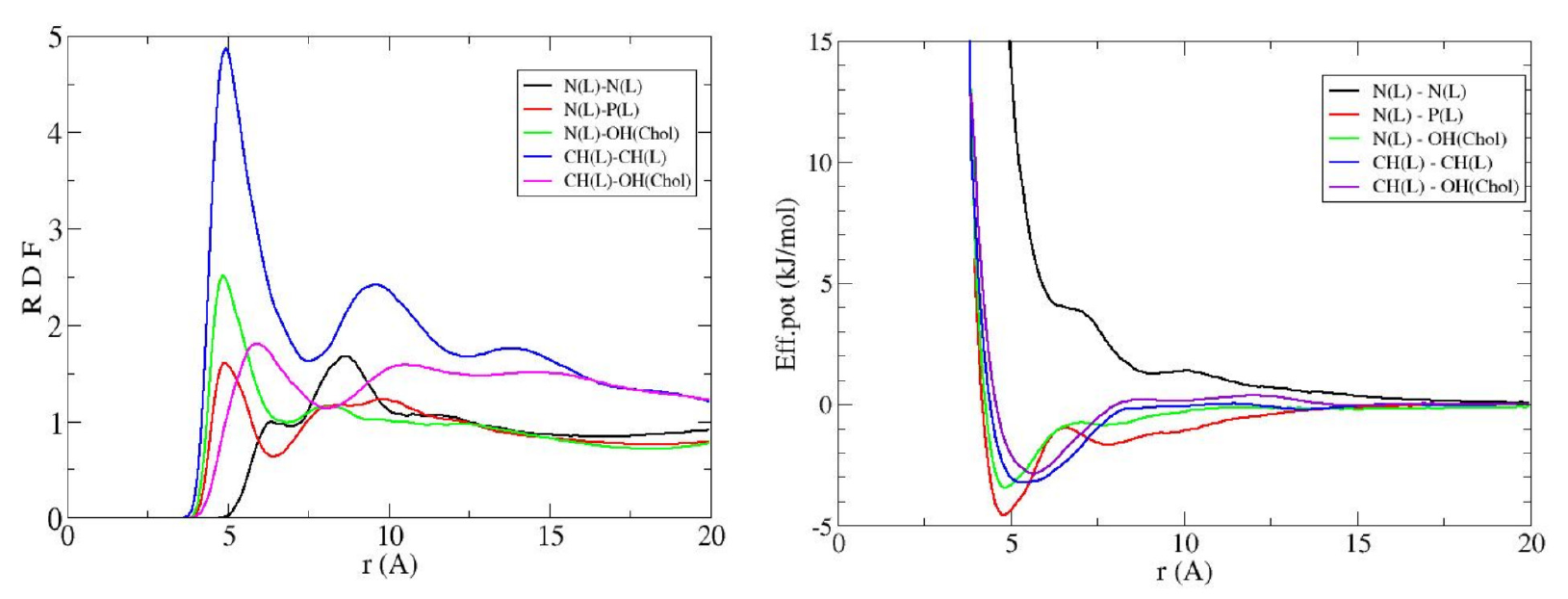}
\caption{Extraction via Inverse Monte Carlo of CG site-site pair potentials from atomistic atomistic RDF's. 5 site-site RDFs (left) and corresponding effective potentials involving DMPC CG sites of total 28 for DMPC-CHOL mixture (right) are shown.}
\label{fig::figD4.4}
\end{figure}

\subsection{Validation of the lipid coarse-grained model}\label{sec::III}

\begin{figure}[]
\centering
\includegraphics[width=0.7\hsize]{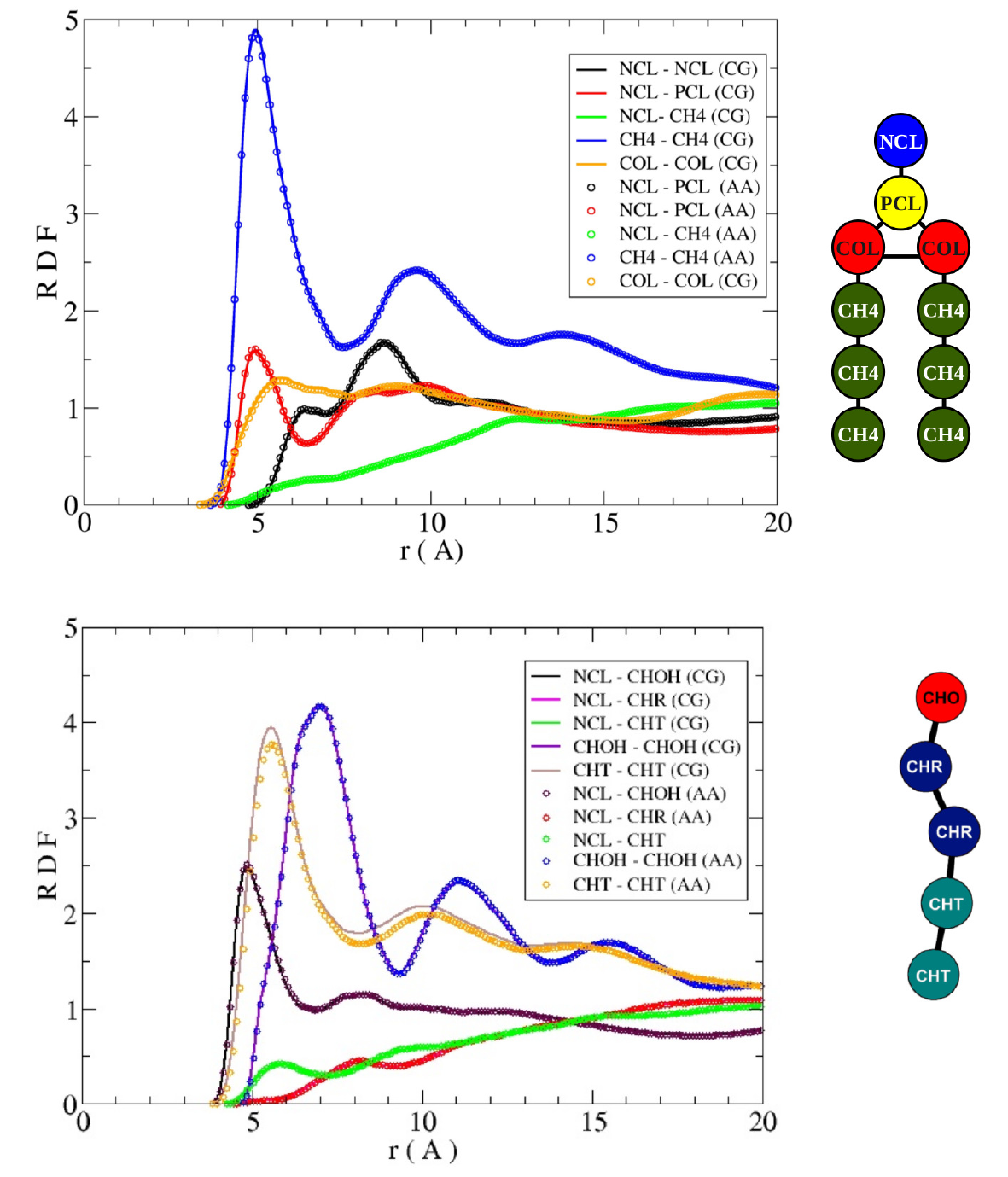}
\caption{Radial distribution functions between different sites of DMPC  (top graph) and Cholesterol (bottom graph) molecules, see site definitions to the right of the graph. Atomistic (points) and coarse-grained (lines) simulations were carried out for a mixture of 30 DMPC, 30 Cholesterol, and 1800 water molecules (CG simulation were without explicit water but in a box of the same size as atomistic simulations including water).}
\label{fig::figD4.5}
\end{figure}

The interaction potentials obtained for the CG models using the inverse Monte Carlo technique
were validated by comparison with atomistic simulations. Figures~\ref{fig::figD4.5} show radial distribution
functions between some selected sites of DMPC lipid and Cholesterol computed in coarse-grained
and atomistic simulations of a mixture of 30 DMPC lipids, 30 cholesterol molecules and 1800 waters.
The result shows a perfect coincidence of the RDFs, which justifies the  approximations made and
the quality of the CG model.

Fig.~\ref{fig::figD4.3} shows
a snapshot of CG DMPC bilayer, which is spontaneously formed in a CG lipid system.
We have carried out a number of simulations of flat lipid bilayers composed of CG lipid models
representing other lipids which can be built from the CG sites presented in Fig. \ref{fig::figD4.2}.
These simulations were carried out  at zero-tension conditions within the atomistic and coarse-grained
models. Table~\ref{tab:comparision} shows
comparison of some properties not related to the radial distribution functions obtained within atomistic
and coarse-grained simulations of a piece of bilayer composed of 128 DMPC lipids. Very good agreement
is observed for the average area per lipid (which is one of the most important parameters for a
lipid bilayer) and for the tail order parameter, and a reasonably good agreement for the bilayer
compressibility. Especially important is the agreement for the order parameter, which shows that
orientational fluctuations of the lipid tails are the same in atomistic and coarse-grained models.

\begin{table}
\begin{tabular}{c c c c c}
\hline
\hline
 & Area per lipid & Compressibility & \multicolumn{2}{ c }{Tail order parameters} \\
 & (\AA) & ($10^{14}$ N/nm) & (1) & (2) \\
\hline
atomistic & 60 & 1.9 & 0.57 & 0.52\\
CG & 59 & 2.5 & 0.56 & 0.52 \\
\hline
\hline
\end{tabular}\caption{Comparison of the properties of the DMPC bilayer obtained from the full atomistic simulations
and the CG model.}
\end{table}\label{tab:comparision}

Table \ref{tab:lip_areas} shows average areas per lipid obtained in coarse-grained simulations carried out in
conditions of zero tension and experiment for a number of lipids. Except DMPC, other lipids included in this table
were not used in the parameterization of the CG potentials. The models for these lipids were built from appropriate sites of DMPC and DOPS lipids shown in Fig. \ref{fig::figD4.2}, and the CG interaction potentials were taken as determined in IMC computations for DMPC and DOPS lipids (some of them shown in Fig. \ref{fig::figD4.4}). One can see generally good agreement with experiment, though simulations show a tendency for some underestimation of the lipid area. The bilayer composed of DSPC lipids was found in the gel phase which again is in agreement with experiment (the temperature of gel phase transition for  DSPC is 55 $^\circ$C). We are not aware on an experimental value of the average lipid area for the gel phase of DSPC, but it is generally accepted that average area per lipid in the gel phase is in the range 43 \AA -- 48 \AA~ for phosphatidylcholine lipids. Also, atomistic simulations of DSPC bilayer in a gel phase \cite{qin09} reports the average lipid area of 44.5 \AA$^2$ which is in good agreement with the result of our coarse-grained model.

\begin{table}
\begin{tabular}{c c c }
\hline
\hline
 Lipid   & \multicolumn{2}{ c }{Area per lipid ($\AA^2$) } \\
       & sim & exp \\
\hline
DMPC  (14:0/14:0 PC)   &    59.0   &   60.5 \cite{kucerka05} \\
SOPC (18:0/18:1n9 PC)  &    60.4   &   61.1 \cite{koenig97} \\
DOPC ( 18:1n9/18:1n9 PC) &    62.0   &   67.4 \cite{kucerka08} \\
DSPC (18:0 /18:0 )     &    43.5$^1$  &   44.5$^{1,2}$ \cite{qin09} \\
\hline
\hline
\end{tabular}\caption{Average areas per lipid. Comparison of simulation results computed in coarse-grained simulations and experiments at T=303K}
$^1$ - bilayer in gel phase
$^2$ - evaluated in atomistic MD simulations
\end{table}\label{tab:lip_areas}
\section{NP and bilayer simulation}\label{sec::NP-bilayer}

Using the methodologies described in sections \ref{sec::NP-P} and \ref{sec::bilayer}
we now can construct a model to simulate the interaction of a DPMC lipid bilayer with a
small hydrophobic NP and a hydrophobic NP associated with one molecule of HSA.
Following is the description of the simulations and the main results.

\subsection{Interaction potentials and parameters of the simulations}

The simulated systems are composed of the NP, the lipids that form the bilayer, the amino
acids of the HSA, and monovalent ions that are used to resemble physiological conditions.
For all interactions we assume two contributions: electrostatic and van der Waals interactions.
For the electrostatic contribution, all charged beads interact through a Coulomb potential between given by:
\begin{equation}\label{eq::coulomb}
U^C(r_{ij}) = \lambda_B k_B T \frac{q_i q_j}{r_{ij}}
\end{equation}
where $r_{ij}$ is the distance between the bead $i$ and the bead $j$, $\lambda_B$ is the
Bjerrum length and $q_i$ and $q_j$ are the the charges of the beads $i$ and $j$ respectively.
The calculation of this long-range interaction was implemented through
a Ewald summation P3M algorithm~\cite{Espresso}. The Bjerrum length is set to 0.71~nm in all cases.

\begin{figure}[]
\centering
\includegraphics[width=0.7\hsize]{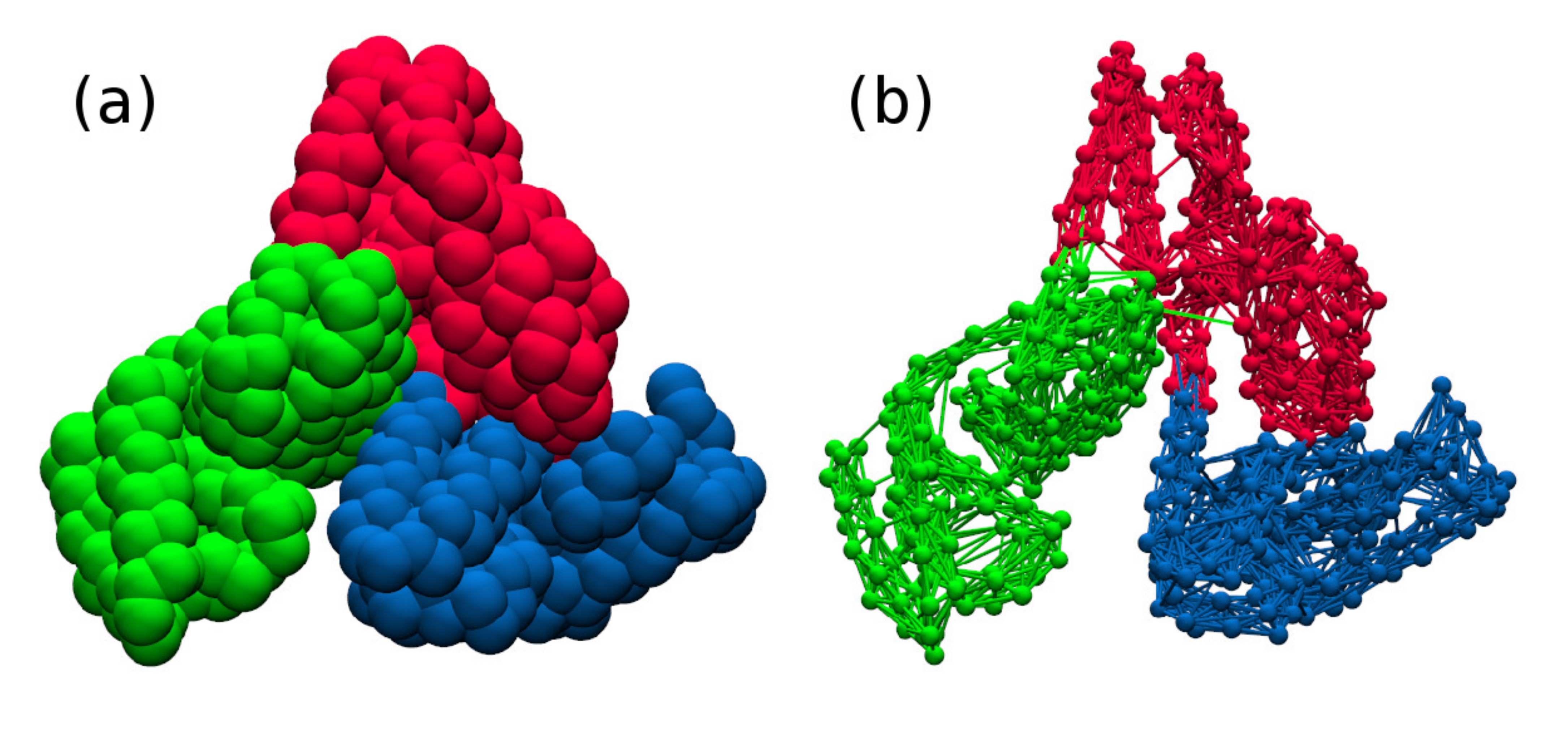}
\caption{(a) CG model of HSA protein used in this study. Each residue is represented by a single bead located at the position of the
$\alpha$-carbon. Each color represents one of the three domains of the HSA molecule. (b) All beads separated by less than
10~nm are connected by stiff harmonic bonds.}
\label{fig::HSA-CG}
\end{figure}

For the van der Waals interactions we use the following model:
\begin{itemize}
	\item Aminoacid -- aminoacid van der Waals interactions: we do not explicitly consider interaction between any pair of
		aminoacids within the single protein molecule as the protein is not allowed to change conformation from the PDB crystal structure.
        To improve the computational
		efficiency instead of simulating the HSA molecule as a rigid body, we connect all residues
		which are separated less than 10~nm by harmonics bonds with a spring constant of $100k_BT$.
		Fig.~\ref{fig::HSA-CG}a shows the resulting CG model for the HSA (build according to PDB ID:1N5U), while
		Fig.~\ref{fig::HSA-CG}b shows the resulting network of bonds (8059 in total).
	\item NP -- aminoacid van der Waals interactions: we used the interaction potential defined
		in Eq.~(\ref{eq::LJRS}) and the same parameters for the aminoacids and the NP obtained
		by the parametrization described in Sec.~\ref{sec::Para}.
	\item Lipid-lipid interactions: the CG models of the lipids in the bilayer and the interactions between the four different types of beads were obtained as described in Sec.~\ref{sec::bilayer}.
	\item Lipid-NP interactions: we used the same interaction potentials as in the case of NP--aminoacid interaction (Eq.~(\ref{eq::LJRS}))
	and assumed that lipid
	beads interact with the surface according to their hydrophobicity. We classify the lipid beads
	into one of two groups: head or tail. The head beads are NCL, PCL and COL (see Fig.~\ref{fig::figD4.2}), and they are considered to be hydrophilic
	with a value of $\epsilon_i=0.1$. The tail beads (labeled CH4 in Fig.~\ref{fig::figD4.2}) are hydrophobic
	and a value of $\epsilon_i=0.75$ is used for these group. The van der Walls radius of all the lipids beads is set
	to $\sigma_i=0.6$~nm.
	\item Lipid-aminoacid interactions: for these interactions we use the same approach as for the NP-residues and NP-lipid. The potential interaction is also based on the hydrophobicity of the beads given by the following modified 12-6 Lennard-Jones potential:
\begin{equation}\label{eq::LJ126}
U_{l,i}(r) =
\left\{
        \begin{array}{ll}
                4\epsilon_{la}\left[\left(\frac{\sigma_{l,i}}{r}\right)^{12}-\left(\frac{\sigma_{l,i}}{r}\right)^{6}\right]
                +\epsilon_{la}(1-\epsilon_{l,i}) & r < r_{c}\\
               4\epsilon_{la}\epsilon_{l,i}\left[\left(\frac{\sigma_{l,i}}{r}\right)^{12}-\left(\frac{\sigma_{l,i}}{r}\right)^{6}\right] & r_c\leq r \leq r_{\mathrm{cut}} \\
                0 & r >  r_{\mathrm{cut}}
        \end{array}
\right.
\end{equation}
where $r$ is the distance from the lipid bed $l$ to the residue $i$,
$\epsilon_{la}$ is a free parameter that scales the interaction energy,
$\epsilon_{l,i}$ is the combined hydrophobicity index of lipid $l$ and the residue $i$ and
is given by $\epsilon_{l,i}=\sqrt{\epsilon_l \epsilon_i}$,
$\sigma_{l,i}$ is the average van der Waals radius of residue $i$ and the lipid $l$, $\sigma_{l,i}=(\sigma_l + \sigma_i)/2$,
$r_c=2^{1/6}\sigma_{l,i}$ and $r_{\mathrm{cut}}$ is the cut-off for the  van der Waals interaction.
As in this work we only study the applicability of the proposed methodology, we do not systematically parameterize the value of $\epsilon_{la}$,
instead we set this scaling parameter to $0.5 k_BT$ for all simulations. This values gives interactions between the lipids and the
residues in the same order of magnitude as the ones reported in~\cite{Bereau2014}.
	\item Ion--ion and ion--molecule interactions: in addition to the Coulomb forces, we include excluded volume interactions by means of a WCA potential. The van der Waals radii of the ions are set to 0.2~nm.
\end{itemize}

For all simulations a NP of radius 2~nm is used with a surface charge of $-0.02$ C/m$^2$.
We use $NVT$ ensemble with the box size was $15 \times 15 \times 20$ nm and we assume
physiological conditions with monovalent salt
concentration of 0.1~M (270 negative and positive ions are placed in the simulation box).
To keep charge neutrality, further 16 positive ions are added.
The bilayer is composed of a total of 762 lipids (381 lipids in each layer).
A Langevin thermostat with a friction coefficient of $\gamma=0.05$ is used and
the units of mass, energy and charge are the same as described in Sec.~\ref{sec::Para}.
The time unit ($\tau$) is obtained by performing a simulation of the bilayer with ions
(no NP or proteins are added) and measuring the lateral diffusion constant.
We obtained a value of $8\times10^{-5}$ nm$^2$/$\tau$, which compared with the experimental value of
5 $\mu^2/\textrm{s}$ gives $\tau=16$ ps.

\subsection{Simulation of a nanoparticle in contact with a lipid bilayer}

\begin{figure}[]
\centering
\includegraphics[width=0.9\hsize]{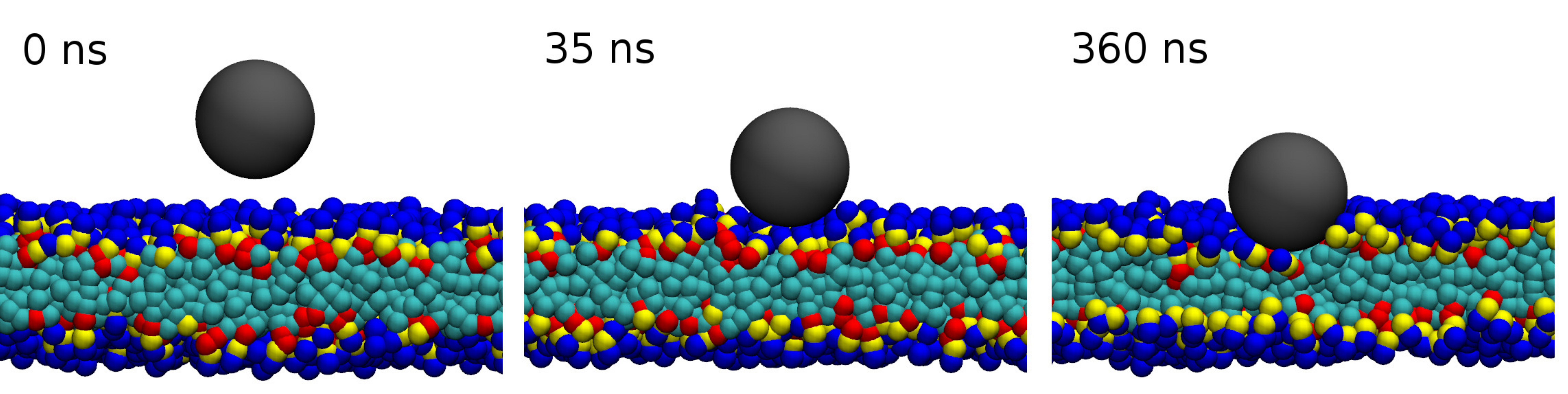}
\caption{Time sequence of the snapshots of the interaction of a DMPC lipid bilayer with a
negative charged hydrophobic NP. The radius of the NP is 2 nm. The ions are not shown.}
\label{fig::bare}
\end{figure}

\begin{figure}[]
\centering
\includegraphics[width=0.9\hsize]{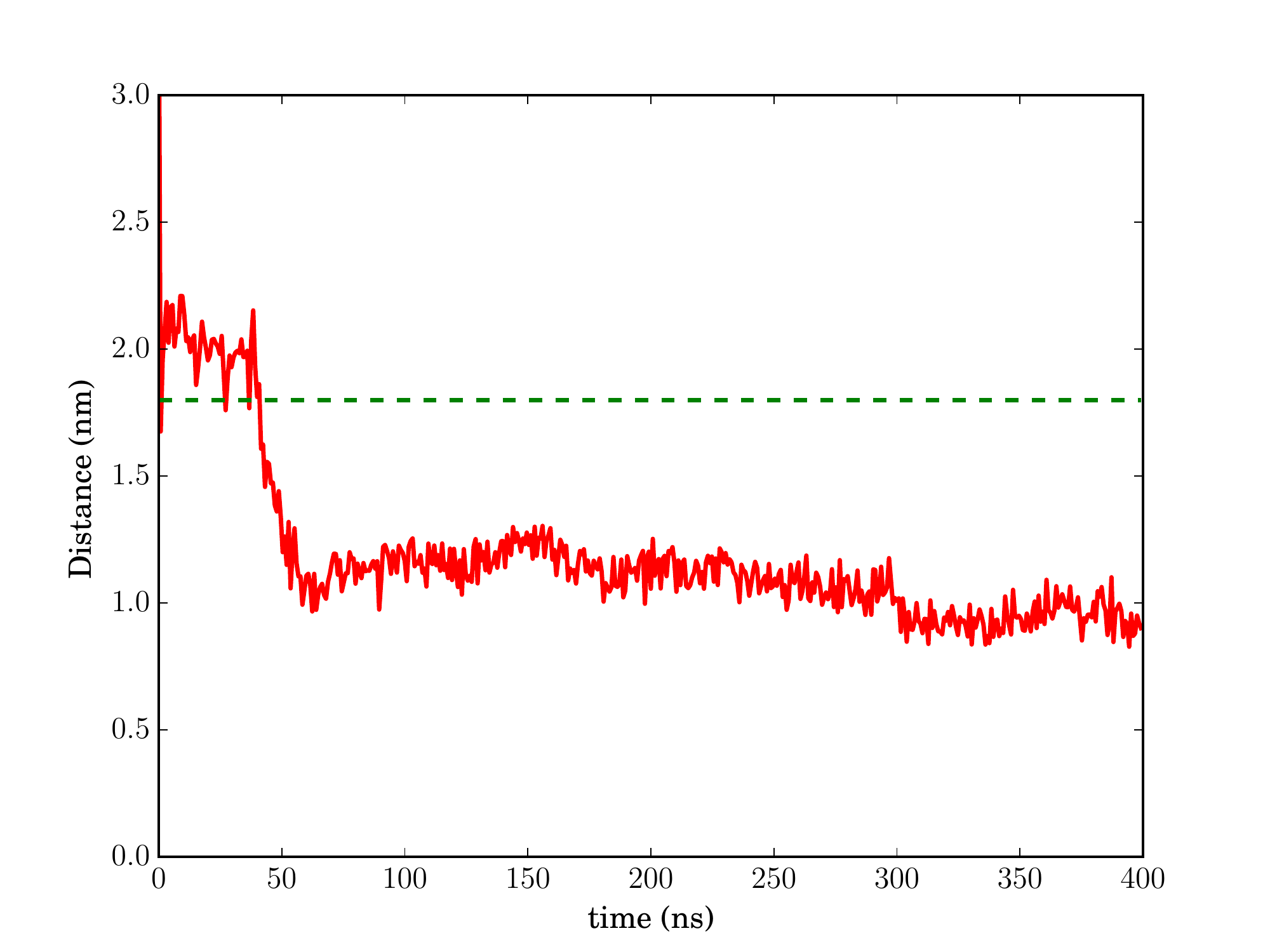}
\caption{Distance of the surface of the bare NP to the center of the bilayer as a function
of time. The dashed line shows the average position of the bilayer surface.}
\label{fig::COMtbare}
\end{figure}

To study the interaction of a bare NP with the lipid bilayer we initially position
the NP close to the bilayer surface and follow the time dynamics of the system.
Three snapshots are shown Fig.~\ref{fig::bare}. From the initial state the NP
adsorbs quickly to the surface of the bilayer and then penetrates to around 1~nm
inside the membrane and stays strongly attached until the end of the simulation
(see snapshots at 35 and 360 ns). To explore the adsorption process in more detail,
the distance of the surface of the NP to the center of the bilayer was recorded
and the results are shown in Fig.~\ref{fig::COMtbare}. The NP reaches the bilayer
in a few nanoseconds and then attaches to the surface for around
50~ns (the dashed line in Fig.~\ref{fig::COMtbare} marks the average position of the surface of the
lipid bilayer, as defined by the position of the maximum of density of the lipid headgroups).
After that, the NP starts penetrating the membrane (in a few nanoseconds).
Then a slow internalization is observed until the NP reaches its final position at approximately 300~ns.
The internalization of the NP is mediated by
the attractions between both type of lipids and the NP. The penetration then stops (or becomes
much slower) because any further displacement requires a substantial change
in the membrane structure. To study long-time dynamics of the system, NP lipid wrapping and uptake
one needs a bigger bilayer or/and NPT ensemble \cite{Lin2010}. Despite of this limitation,
the results obtained with our methodology agree with a
recent report~\cite{Ding20148703} for the absorption of a hydrophobic NP with a
membrane composed of lipids and specialized receptors.

\subsection{Simulation of a nanoparticle-protein complex complex in contact with a lipid bilayer}

\begin{figure}[]
\centering
\includegraphics[width=0.9\hsize]{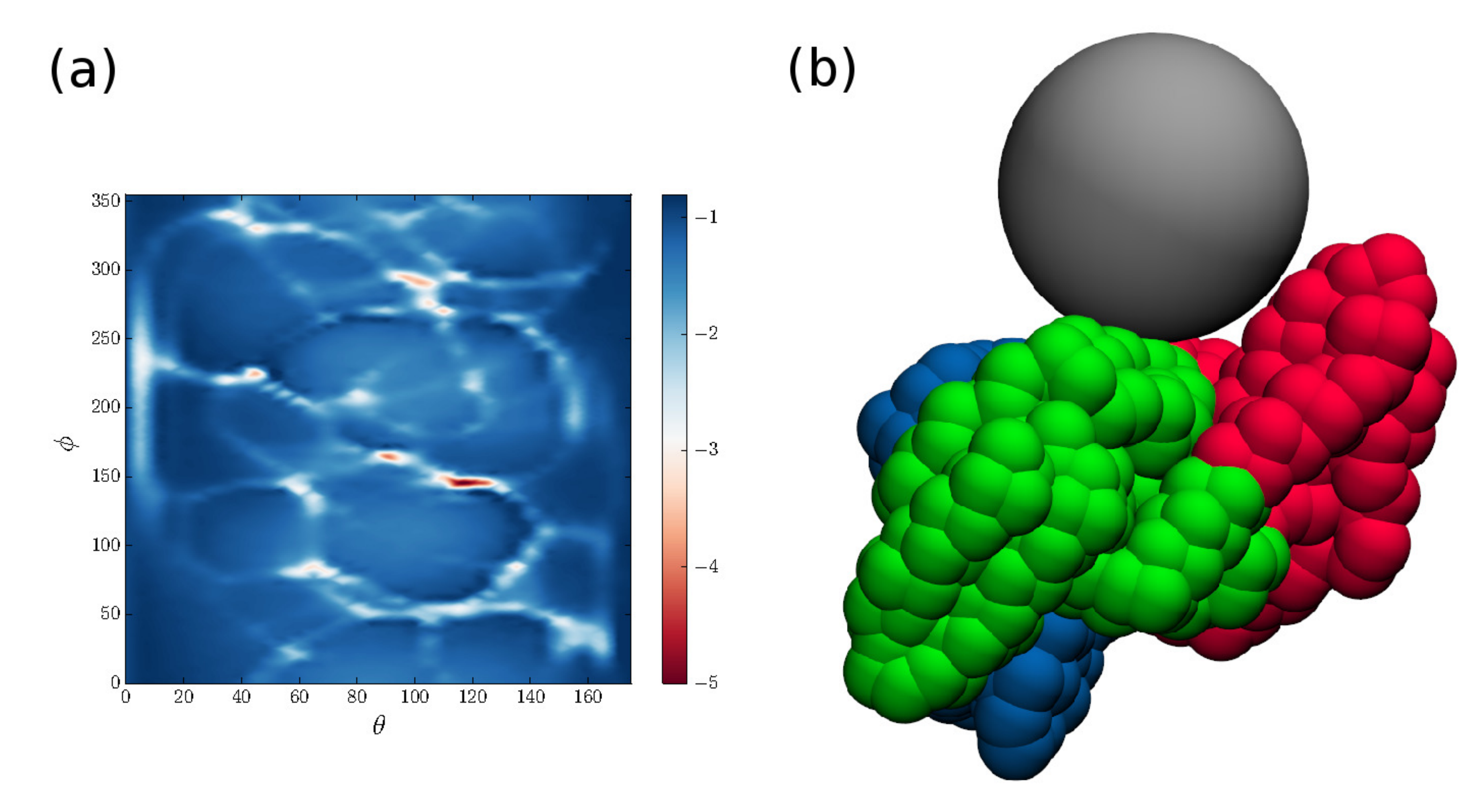}
\caption{(a) Initial state of the NP-HSA complex. (b) Surface map of the
adsorption orientations of HSA onto a 2~nm negatively charged hydrophobic NP.}
\label{fig::NPHSAmap}
\end{figure}

It is well accepted that in more realistic situations, the NP before
it reaches the cell membrane gets coated by proteins and that
this NP-protein complex is responsible for the final fate of the
NP~\cite{MASD2012}. Considering this, we now simulate the interaction of a NP-protein
complex, where protein is represented by a single HSA molecule.
As shown in Sec.~\ref{sec::NP-P}, not all orientations in which protein adsorbs
onto a surface are equally probable and for our simulation of the interaction
of a hydrophobic NP with a DPMC lipid membrane we first calculate the adsorption
energy map of HSA onto a 2~nm of radius hydrophobic NP. The adsorption
map is shown in Fig.~\ref{fig::NPHSAmap}a. We can see that, as in the cases discussed above, the
energy landscape contains more than one minimum. We found the average adsorption energy of $-1.7$ $k_BT$
and as the initial orientation for the simulation we selected the
orientation $\theta,\phi=145^{\circ},110^{\circ}$), which corresponds to adsorption
energy of $-2.7$ $k_BT$ and the corresponding
complex NP-HSA is shown in Fig.~\ref{fig::NPHSAmap}b.

\begin{figure}[]
\centering
\includegraphics[width=0.9\hsize]{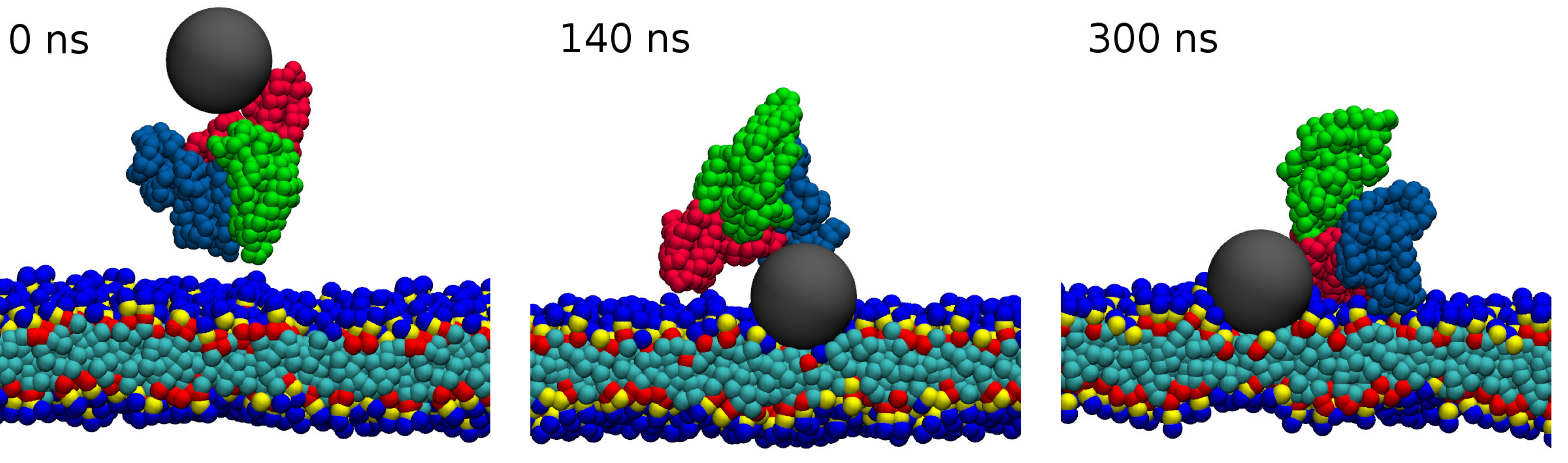}
\caption{Time sequence of the snapshots of the interaction of a DMPC lipid bilayer with a
negatively charged hydrophobic NP complex with one HSA molecule. The radius of the NP is 2 nm.
 The ions are not shown.}
\label{fig::npcorona}
\end{figure}

\begin{figure}[]
\centering
\includegraphics[width=0.9\hsize]{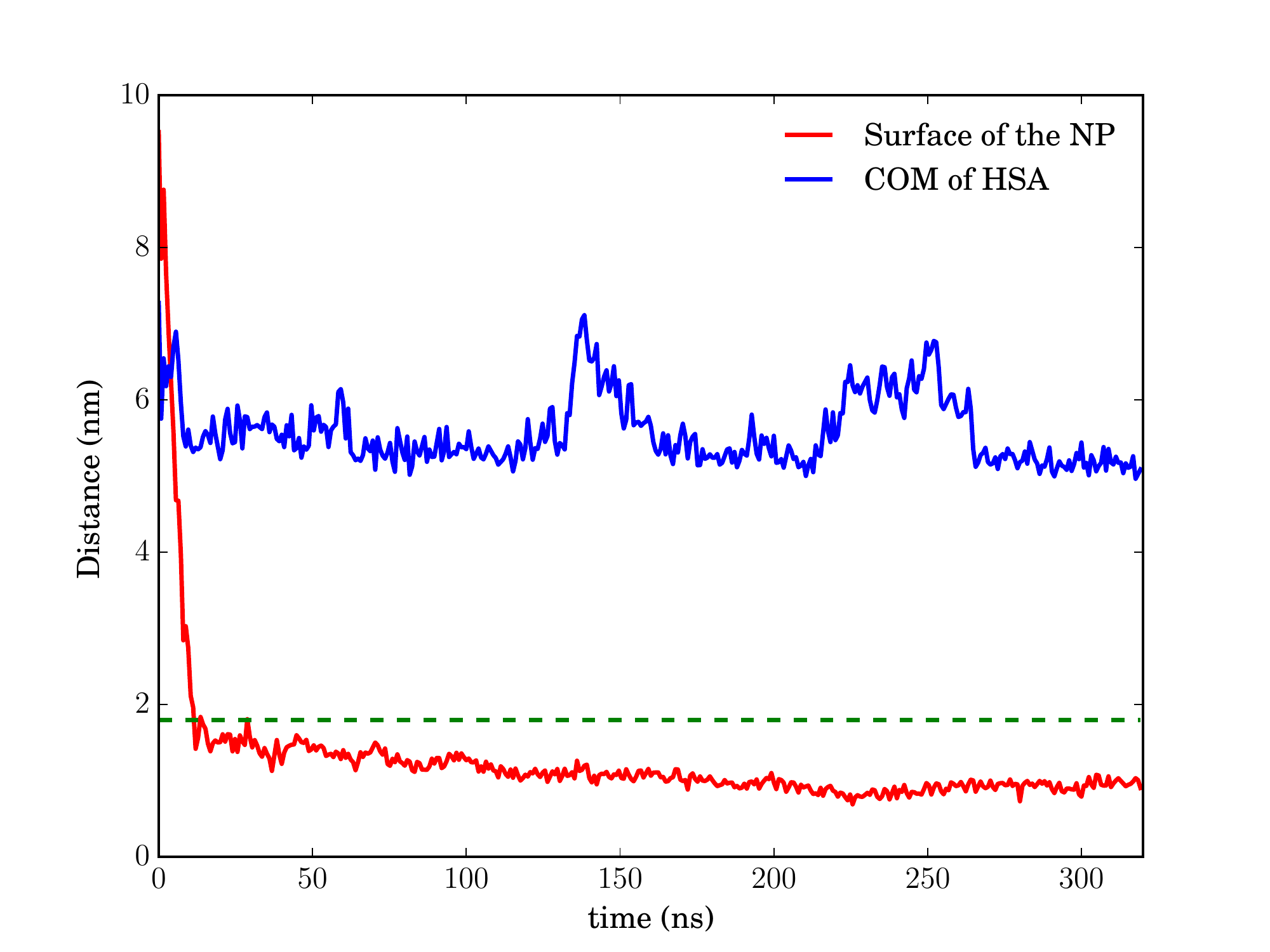}
\caption{Distance of the surface of the complex of NP with one HSA to the center of the bilayer
and the COM of the HSA as a function of time.
The dashed line shows the average position of the bilayer surface.}
\label{fig::COMtNPcorona}
\end{figure}

Figure~\ref{fig::npcorona} shows a sequence of snapshots from simulation
of the NP-HSA complex with the lipid bilayer. In the initial state, the protein is
facing the bilayer. In the simulation, the HSA at first moves in front of the membrane and prevents a direct
contact between the  NP and the lipids. Then, the NP-HSA complex rotates so that the NP faces the bilayer and
starts penetrating the membrane. Fig.~\ref{fig::COMtNPcorona}
shows the distance of the NP surface to the center of the membrane. The rotation is reflected in the sudden
change of the position of COM of the HSA. After this quick rearrangement the NP starts the penetration while the
protein stays attached to the NP for the whole simulation but moves around
the surface of the NP as can be seen in the snapshot for the times 140 and 300 ns in Fig.~\ref{fig::npcorona}.
This movement of the HSA molecule can also be observed from the curve of the COM of the HSA curves
as a function of time (Fig.~\ref{fig::COMtNPcorona}).

Comparing the two simulations we see that the presence of the HSA dramatically changes
the interaction of the NP with the membrane. We can envision that a NP, fully coated NP with HSA,  will not be able
to penetrate into the membrane, so the coating is changing completely its biological reactivity.

\section{Conclusions}\label{sec::conclusions}

In this work, we presented a multiscale methodology for modelling interactions at bionano interface,
which is central for understanding uptake and toxicity of nanomaterials. We used systematic coarse-graining
techniques to reduced the complexity of the problem by removing some degrees of freedom and
focussing on the properties of interest. Since the CG models consists
of about ten times less interaction centres than the atomistic model, and the solvent (water) is
not modeled explicitly, simulations of the CG model take 2-3 order less CPU time compared with
atomistic simulations for equal system size, or, alternatively, CG model can be used for simulations
of whole proteins, small NPs  and sufficiently large cell membrane fragments at the scale of tens of nanometers.
We have parameterised and validated the model against experiments.

The technique for coarse-graining NP-protein interaction, which we presented in Section \ref{sec::NP-P} can be used to calculate the binding energies for
arbitrary plasma, cytosolic or membrane proteins, rank them by binding affinity to the NP and predict the content
of NP protein corona. Our calculations show that the NP surface charge has a small effect on the adsorption
energies in comparison to van der Waals interactions between the residues and the surface. We also find that
the charge of the NP does not influence much the orientation, in which the protein prefer to adsorb.
On the other hand, we have shown the size of the NP has a big effect on the adsorption energy maps, as the
curvature of the NP determine the sections of the protein that can interact with the surface.
Based on our simulations results, we can predict bigger proteins adsorb stronger on the inorganic surfaces, even
for small NPs, in agreement with the Vroman effect. We have also demonstrated that a rigid protein model is justified at least
for small globular proteins. In Section \ref{sec::bilayer}, we have parameterised a CG lipid and cholesterol model,
which reproduce the key bilayer properties of atomistic model of the same system. Finally, in Section \ref{sec::NP-bilayer}, we have shown how the CG
lipid and NP-protein models can be combined to model NP-cell membrane interactions and NP attachment and uptake.

%
\bibliographystyle{unsrt}
\bibliography{refs}

\begin{thebibliography}{10}

\bibitem{1-7}
S.~Sharifi, S.~Behzadi, S.~Laurent, M.L. Forrest, P.~Stroevee, and M.~Mahmoudi.
\newblock Toxicity of nanomaterials.
\newblock {\em Chem. Soc. Rev.}, 41:2323, 2012.

\bibitem{1-3}
P.~J.~A. Borm and et~al.
\newblock The potential risks of nanomaterials.
\newblock {\em Part. Fibre Toxicol.}, 3:11, 2006.

\bibitem{1-8}
H.J. Johnston, G.R. Hutchison, F.M. Christensen, and et~al.
\newblock Identification of the mechanisms that drive the toxicity of tio2
  particulates: the contribution of physicochemical characteristics.
\newblock {\em Part. Fibre Toxicol.}, 6:33, 2009.

\bibitem{1-9}
A.~E. Nel, L.~Maedler, D.~Velegol, and et~al.
\newblock Understanding biophysicochemical interactions at the nano-bio
  interface.
\newblock {\em Nat. Mater.}, 8:543, 2009.

\bibitem{1-10}
A.~Verma, O.~Uzun, Y.~Hu, and et~al.
\newblock Surface-structure-regulated cell-membrane penetration by
  monolayer-protected nanoparticles.
\newblock {\em Nat. Mater.}, 7:588, 2008.

\bibitem{1-11}
T.~Schlick, R.~Collepardo-Guevara, L.A. Halvorsen, and et~al.
\newblock Biomolecular modeling and simulation: a field coming of age.
\newblock {\em Quarterly Rev. Biophys.}, 44:191, 2011.

\bibitem{1-12}
L.G. Valerio~Jr.
\newblock In silico toxicology for the pharmaceutical sciences.
\newblock {\em Toxicol. Appl. Pharmacol.}, 241:356, 2009.

\bibitem{1-13}
F.~Nigsch, N.J. Macaluso, J.~B. Mitchell, and D.~Zmuidinavicius.
\newblock Computational toxicology: an overview of the sources of data and of
  modelling methods.
\newblock {\em Expert Opin. Drug Metab. Toxicol.}, 5:1, 2009.

\bibitem{1-14}
J.C. Dearden.
\newblock In silico prediction of drug toxicity.
\newblock {\em J. Comput.-Aided Mol. Des.}, 17:119, 2003.

\bibitem{1-54}
A.~P. Lyubartsev and A.~L. Rabinovich.
\newblock Recent development in computer simulations of lipid bilayers.
\newblock {\em Soft Matter}, 7:25, 2011.

\bibitem{1-58}
J.~Wong-Ekkabut, S.~Baoukina, W.~Triampo, I.-M. Tang, and D.~P. Tieleman.
\newblock Computer simulation study of fullerene translocation through lipid
  membranes.
\newblock {\em Nat. Nanotechnol.}, 3:363, 2008.

\bibitem{1-58a}
W.C. Hou, B.~Y. Moghadam, P.~Westerhoff, and J.~D. Posner.
\newblock Distribution of fullerene nanomaterials between water and model
  biological membranes.
\newblock {\em Langmuir}, 27:11899, 2011.

\bibitem{1-58b}
K.~Yang and Y.~Q. Ma.
\newblock Computer simulation of the translocation of nanoparticles with
  different shapes across a lipid bilayer.
\newblock {\em Nat. Nanotechnol.}, 5:579, 2010.

\bibitem{1-58c}
L.~Monticelli, E.~Salonen, P.C. Ke, and I.~Vattulainen.
\newblock Effects of carbon nanoparticles on lipid membranes: a molecular
  simulation perspective.
\newblock {\em Soft Matter}, 5:4433, 2009.

\bibitem{1-60}
S.~Izvekov and G.A. Voth.
\newblock Multiscale coarse-graining method for biomolecular systems.
\newblock {\em J. Phys. Chem. B}, 109:2469, 2005.

\bibitem{1-60a}
G.~S. Ayton, W.~G. Noid, and G.~A. Voth.
\newblock Multiscale modeling of biomolecular systems: in serial and in
  parallel.
\newblock {\em Curr. Opin. Struct. Biol.}, 17:192, 2007.

\bibitem{1-61}
A.~P. Lyubartsev and A.~Laaksonen.
\newblock Calculation of effective interaction potentials from radial
  distribution functions: A reverse monte carlo approach.
\newblock {\em Phys. Rev. E}, 52:3730, 1995.

\bibitem{1-62}
A.~P. Lyubartsev, A.~Mirzoev, L.-J. Chen, and A.~Laaksonen.
\newblock Systematic coarse-graining of molecular models by the newton
  inversion method.
\newblock {\em Faraday Discuss.}, 144:43, 2010.

\bibitem{1-65}
A.~P. Lyubartsev and A.~Laaksonen.
\newblock Effective potentials for ion-dna interactions.
\newblock {\em J. Chem. Phys.}, 111:11207, 1999.

\bibitem{Toz2005}
V.~Tozzini.
\newblock {Coarse-grained models for proteins}.
\newblock {\em {Curr. Opin. Struct. Biol.}}, {15}({2}):{144--150}, {2005}.

\bibitem{BerDes2009}
T.~Bereau and M.~Deserno.
\newblock {Generic coarse-grained model for protein folding and aggregation}.
\newblock {\em {J. Chem. Phys.}}, {130}({23}):{235106}, {2009}.

\bibitem{Tak2012}
S.~Takada.
\newblock {Coarse-grained molecular simulations of large biomolecules}.
\newblock {\em {Curr. Opin. Struct. Biol.}}, {22}({2}):{130--137}, {2012}.

\bibitem{WeiKno2013}
S.~Wei and T.~Knotts.
\newblock {A coarse grain model for protein-surface interactions}.
\newblock {\em {J. Chem. Phys.}}, {139}({9}):095102, {2013}.

\bibitem{1-64}
V.~Lobaskin, A.~P. Lyubartsev, and P.~Linse.
\newblock Effective macroion-macroion potentials in asymmetric electrolytes.
\newblock {\em Phys. Rev. E}, 63:020401, 2001.

\bibitem{1-67}
M.~Brunner, C.~Bechinger, W.~Strepp, V.~Lobaskin, and H.~H. von Gruenberg.
\newblock Density-dependent pair interactions in 2d colloidal dispersions.
\newblock {\em Europhys. Lett.}, 58:926, 2002.

\bibitem{1-71}
I.~Lynch, A.~Salvati, and K.~A. Dawson.
\newblock Protein-nanoparticle interactions. what does the cell see?
\newblock {\em Nat. Nanotechnol.}, 4:546, 2009.

\bibitem{1-72}
I.~Lynch, K.~A. Dawson, and S.~Linse.
\newblock Detecting cryptic epitopes created by nanoparticles.
\newblock {\em Sci. STKE2006}, page pe14, 2006.

\bibitem{1-72a}
T.~Cedervall and et~al.
\newblock Understanding the nanoparticle protein corona using methods to
  quantify exchange rates and affinities of proteins for nanoparticles.
\newblock {\em Proc. Natl. Acad. Sci. U.S.A.}, 104:2050, 2007.

\bibitem{1-72b}
S.~Lindman and et~al.
\newblock Systematic investigation of the thermodynamics of hsa adsorption to
  n-iso-propylacrylamide/n-tert-butylacrylamide copolymer nanoparticles.
  effects of particle size and hydrophobicity.
\newblock {\em Nanoletters}, 7:914, 2007.

\bibitem{1-72c}
L.~T. Allen and et~al.
\newblock Surface-induced changes in protein adsorption and implications for
  cellular phenotypic responses to surface interaction.
\newblock {\em Biomaterials}, 27:3096, 2006.

\bibitem{1-85}
C.~E. Radke and J.~M. Prausnitz.
\newblock Thermodynamics of multisolute adsorption from dilute liquid
  solutions.
\newblock {\em AIChE J.}, 18:761, 1972.

\bibitem{1-36}
A.~Lesniak, A.~Campbell, M.~P. Monopoli, I.~Lynch, A.~Salvati, and K.~A.
  Dawson.
\newblock Serum heat inactivation affects protein corona composition and
  nanoparticle uptake.
\newblock {\em Biomaterials}, 31:9511, 2010.

\bibitem{Noi2013}
W.~G. Noid.
\newblock Perspective: Coarse-grained models for biomolecular systems.
\newblock {\em J. Chem. Phys.}, 139(9):090901, 2013.

\bibitem{Lopez2015}
H.~Lopez and V.~Lobaskin.
\newblock Coarse-grained model of adsorption of blood plasma proteins onto
  nanoparticles.
\newblock {\em arxiv:1508.01450}, 2015.

\bibitem{MiyJer1996}
S.~Miyazawa and R.L. Jernigan.
\newblock {Residue-residue potentials with a favorable contact pair term and an
  unfavorable high packing density term, for simulation and threading}.
\newblock {\em {J. Mol. Biol.}}, {256}({3}):{623--644}, {1996}.

\bibitem{KTMH2008}
Y.~Kim, C.~Tang, G.~Clore, and G.~Hummer.
\newblock {Replica exchange simulations of transient encounter complexes in
  protein-protein association}.
\newblock {\em {Proc. Natl. Acad. Sci. USA}}, {105}({35}):{12855--12860},
  {2008}.

\bibitem{KimHum2008}
Y.~Kim and G.~Hummer.
\newblock {Coarse-grained models for simulations of multiprotein complexes:
  application to ubiquitin binding}.
\newblock {\em {J. Mol. Biol.}}, {375}({5}):{1416--1433}, {2008}.

\bibitem{ARS2005}
M.~Agashe, V.~Raut, S.~Stuart, and R.~Latour.
\newblock Molecular simulation to characterize the adsorption behavior of a
  fibrinogen $\gamma$-chain fragment.
\newblock {\em Langmuir}, 21(3):1103--1117, 2005.

\bibitem{SWL2005}
Y.~Sun, W.~Welsh, and R.~Latour.
\newblock {Prediction of the orientations of adsorbed protein using an
  empirical energy function with implicit solvation}.
\newblock {\em {Langmuir}}, {21}({12}):{5616--5626}, {2005}.

\bibitem{Dariaetal2010}
D.~Kokh, S.~Corni, P.~Winn, M.~Hoefling, K.~Gottschalk, and R.~Wade.
\newblock Prometcs: An atomistic force field for modeling protein−metal
  surface interactions in a continuum aqueous solvent.
\newblock {\em J. Chem. Theory Comput.}, 6(5):1753--1768, 2010.

\bibitem{Espresso}
H.~Limbach, A.~Arnold, B.~Mann, and C.~Holm.
\newblock {ESPResSo - an extensible simulation package for research on soft
  matter systems}.
\newblock {\em {Comput. Phys. Commun.}}, {174}({9}):{704--727}, {2006}.

\bibitem{CHL2003}
W.~Chen, H.~Huang, C.~Lin, F.~Lin, and Y.~Chan.
\newblock Effect of temperature on hydrophobic interaction between proteins and
  hydrophobic adsorbents: Studies by isothermal titration calorimetry and the
  van't hoff equation.
\newblock {\em Langmuir}, 19(22):9395--9403, 2003.

\bibitem{Petal2010}
S.~Lacerda, Jung~J. Park, C.~Meuse, D.~Pristinski, M.~Becker, A.~Karim, and
  J.~Douglas.
\newblock Interaction of gold nanoparticles with common human blood proteins.
\newblock {\em ACS Nano}, 4(1):365--379, 2010.

\bibitem{VDF2013}
P.~Vilaseca, K.~Dawson, and G.~Franzese.
\newblock Understanding and modulating the competitive surface-adsorption of
  proteins through coarse-grained molecular dynamics simulations.
\newblock {\em Soft Matter}, 9:6978--6985, 2013.

\bibitem{Vijay-Kumar1987}
S.~Vijay-Kumar, C.~Bugg, and W.~Cook.
\newblock Structure of ubiquitin refined at 1.8 \aa resolution.
\newblock {\em J. Mol. Biol.}, 194:531--544, 1987.

\bibitem{Momma2011}
K.~Momma and F.~Izumi.
\newblock {\it VESTA3} for three-dimensional visualization of crystal,
  volumetric and morphology data.
\newblock {\em Journal of Applied Crystallography}, 44:1272--1276, 2011.

\bibitem{Brandt2014}
E.~G. Brandt and A.~Lyubartsev.
\newblock Systematic optimization of a force field for classical simulations of
  interfaces between tio$_2$ surfaces and water.
\newblock In preparation.

\bibitem{Berendsen1984}
H.~J.~C. Berendsen, J.~P.~M. Postma, W.~F. van Gunsteren, A.~DiNola, and J.~R.
  Haak.
\newblock Molecular dynamics with coupling to an external bath.
\newblock {\em J. Chem. Phys.}, 81:3684--3690, 1984.

\bibitem{Hoover1985}
W.~Hoover.
\newblock Canonical dynamics: Equilibrium phase-space distributions.
\newblock {\em Phys. Rev. A}, 31:1695--1697, 1985.

\bibitem{Nose1984}
S.~Nose.
\newblock A unified formulation of the constant temperature molecular dynamics
  methods.
\newblock {\em J. Chem. Phys.}, 81:511--519, 1984.

\bibitem{JoaLyu2012}
J.~P.~M. J\"ambeck and A.~P. Lyubartsev.
\newblock Derivation and systematic validation of a refined all-atom force
  field for phosphatidylcholine lipids.
\newblock {\em J. Phys. Chem. B}, 116(10):3164--3179, 2012.

\bibitem{JoaLyu2013}
J.~P.~M. J\"ambeck and A.~P. Lyubartsev.
\newblock Another piece of the membrane puzzle: Extending slipids further.
\newblock {\em J. Chem. Theory Comput.}, 9(1):774--784, 2013.

\bibitem{MirLyu2013}
A.~Mirzoev and A.~P. Lyubartsev.
\newblock Magic: Software package for multiscale modeling.
\newblock {\em J. Chem. Theory Comput.}, 9(3):1512--1520, 2013.

\bibitem{qin09}
{S.-S.} Qin, Z.~W. Yu, and {Y.-X.} Yu.
\newblock Structural characterization on the gel to liquid-crystal phase
  transition of fully hydrated {DSPC} and {DSPE} bilayers.
\newblock {\em J. Phys. Chem. B}, 113:8114 -- 8123, 2009.

\bibitem{kucerka05}
N.~Ku{\v c}erka, Y.~Liu, N.~Chu, H.~I. Petrache, S.~Tristram-Nagle, and J.~F.
  Nagle.
\newblock Structure of fully hydrated fluid phase {DMPC} and {DLPC} lipid
  bilayers using {X-ray} scattering from oriented multilamellar arrays and from
  unilamellar vesicles.
\newblock {\em Biophys. J.}, 88:2626 -- 2637, 2005.

\bibitem{koenig97}
B.~W. K{\"o}enig, H.~H. Strey, and K.~Gawrisch.
\newblock Membrane lateral compressibility determined by {NMR} and {X}-ray
  diffraction: effect of acyc chain polyunsaturation.
\newblock {\em Biophys. J.}, 73(4):1954 -- 1966, 1997.

\bibitem{kucerka08}
N.~Ku{\v c}erka, J.~F. Nagle, J.~N. Sachs, S.~E. Feller, J.~Pencer, A.~Jackson,
  and J.~Katsaras.
\newblock Lipid bilayer structure determined by the simultaneous analysis of
  neutron and {X-Ray} scattering data.
\newblock {\em Biophys. J.}, 95(5):2356 -- 2367, 2008.

\bibitem{Bereau2014}
T.~Bereau, Z.-J. Wang, and M.~Deserno.
\newblock More than the sum of its parts: Coarse-grained peptide-lipid
  interactions from a simple cross-parametrization.
\newblock {\em J. Chem. Phys.}, 140(11):115101, 2014.

\bibitem{Lin2010}
J.~Lin, H.~Zhang, Z.~Chen, and Y.~Zheng.
\newblock Penetration of lipid membranes by gold nanoparticles: Insights into
  cellular uptake, cytotoxicity, and their relationship.
\newblock {\em ACS Nano}, 4(9):5421--5429, 2010.

\bibitem{Ding20148703}
Hong-Ming D. and Yu-Qiang M.
\newblock Computer simulation of the role of protein corona in cellular
  delivery of nanoparticles.
\newblock {\em Biomaterials}, 35(30):8703 -- 8710, 2014.

\bibitem{MASD2012}
M.~Monopoli, C.~Aberg, A.~Salvati, and Kenneth~A. Dawson.
\newblock {Biomolecular coronas provide the biological identity of nanosized
  materials}.
\newblock {\em {Nat. Nanotechnol.}}, {7}({12}):{779--786}, {2012}.

\end{thebibliography}

\end{document}